\documentclass[12pt]{article}
\usepackage[bottom]{footmisc}
\usepackage{dcolumn}  %% For STARGAZER
\usepackage{amsfonts}
\usepackage{bm}
\usepackage{url}
\usepackage{comment}
\usepackage{nccmath, amsmath, amssymb}
\usepackage{graphicx}
\usepackage[top=30truemm,bottom=30truemm,left=25truemm,right=25truemm]{geometry}

\newcommand{\E}{\mathrm{E}}

\usepackage{changepage}
\usepackage{ragged2e} %折り返し

\usepackage[english]{babel}
\usepackage{natbib} %%Author + year style

\usepackage{setspace}

\newdimen\dummy
\dummy=\oddsidemargin
\begin{document}

\begin{flushleft}
{\Large {\bf The Resilience of FDI to Natural Disasters through Industrial Linkages}}
\end{flushleft}

\vspace{0.3cm}

\noindent
{\bf Hayato Kato\footnote{Graduate School of Economics, Osaka University, 1-7 Machikaneyama, Toyonaka, Osaka, 560-0043, Japan. {\it E-mail address:} \ hayato.kato@econ.osaka-u.ac.jp} \ $\cdot$ \ Toshihiro Okubo\footnote{Faculty of Economics, Keio University, 2-15-45 Mita, Minato-ku, Tokyo, 108-8345, Japan. {\it E-mail address:} \ okubo@econ.keio.ac.jp}}

\

\

\

\noindent
{\bf Abstract} \ 
\noindent When do multinationals show resilience during natural disasters?
To answer this, we develop a simple model in which foreign multinationals and local firms in the host country are interacted through input-output linkages.
When natural disasters seriously hit local firms and thus increase the cost of sourcing local intermediate inputs, most multinationals may leave the host country.
However, they are likely to stay if they are tightly linked with local suppliers and face low trade costs of importing foreign intermediates.
We further provide a number of extensions of the basic model to incorporate, for example, multinationals with heterogeneous productivity and disaster reconstruction.

\vspace{0.3cm}


\vspace{0.3cm}


\vspace{0.3cm}

\begin{spacing}{1.0}
\noindent
\hrulefill \hspace{10cm} \\
\noindent 
{\footnotesize This is a substantially revised version of our earlier working paper (\citealp{KatoOkubo2017}).
It is conducted as a part of the Project ``Economic Policy Issues in the Global Economy'' undertaken at the Research Institute of Economy, Trade and Industry (RIETI).
We would like to thank Co-Editor, Eric Strobl, and two anonymous referees for helpful suggestions.
Thanks also go to Akira Sasahara for extensive discussions and conference/seminar participants at RIETI, Hosei U, Osaka U, Kobe U and EWMES2021 for useful comments.
Financial support from the the Japan Society for the Promotion of Science (Grant Numbers: JP19K13693; JP99K13693; JP20H01495) are gratefully acknowledged.
All remaining errors are our sole responsibility.}
\end{spacing}

\pagebreak

\section{Introduction}

Multinational enterprises (MNEs) play a vital role in helping developing countries grow by contributing to local employment and productivity improvement in normal times.
However, MNEs may play a more crucial role during crisis times, especially in natural disasters.
Thailand's experience of large-scale floods in 2011 is a notable example of the destructive effects of a natural disaster on multinationals.
The estimated economic damage is 46.5 billion USD across all sectors, and manufacturing alone suffered 32 billion USD of damage (\citealp{AbeYe2013}; \citealp{HaraguchiLall2015}).
Seven industrial parks and 904 factories were inundated, more than half of which were Japanese MNEs.
Thus, the business operation of many factories completely ceased for one to two months.
Even non-inundated multinational plants were forced to reduce production due to the lack of parts from damaged suppliers. 
The disaster-hit economy hopes multinationals to show resilience so that they continue to source local intermediates and help local industry recover.
The Bank of Thailand cautioned against this optimistic idea by stating that, unless the government took appropriate measures, the flood would relocate more multinationals to other Asian countries in the long run (\citealp{BOT2012}).
The exit of footloose multinationals would weaken their input-output linkages with local suppliers, damaging the host economy further.
Indeed, using country-level data, \cite{EscalerasRegister2011} and \cite{Doytch2020} find that natural disasters have negative effects on aggregate foreign direct investment (FDI) inflows.\footnote{\cite{Anuchitworawong2015} construct a severity index of natural disasters consisting of their frequencies and damages and find that this negatively affects aggregate FDI inflows into Thailand in 1971--2012.
See also \cite{Batalaetal2021}; and \cite{Neiseetal2021} for recent studies using aggregate level FDI.}

Only a limited attention in the literature has been paid to how MNEs respond to natural disasters and little is known about underlying mechanisms, despite its importance in the real world. 
This paper theoretically investigates under what conditions multinationals leave their host country once it is hit by negative supply shocks and which factor helps them stay there.
To address these, our model takes into account two noticeable characteristics of multinationals: (i) input-output (or vertical) industrial linkages and (ii) footloose-ness.
Vertical linkages imply that multinationals have  complementarities with local industry (\citealp{MarkusenVenables1999}).
Sourcing by multinationals helps the local supplying industry grow, leading to a lower price of inputs and thus benefiting themselves.
The footloose-ness arises because the location choice of foreign capital is based on the comparison of profits made in different countries.
Multinationals are sensitive to negative shocks in their location; compared with local firms, they are likely to enter and exit a host country frequently (\citealp{GorgStrobl2003}; \citealp{BernardJensen2007}).
We model a natural disaster as an increase in the fixed cost for local suppliers, which can be interpreted as a damage to their plants.
This modeling captures an indirect damage to multinationals through their linkages with damaged local suppliers, as highlighted in the 2011 Thailand floods.

Our main findings are threefold.
First, multiple equilibria, one in which multinationals enter and the other in which they do not, may exist.
Second, a substantial damage to the local suppliers' fixed cost results in a switch from one equilibrium with multinationals to the other without them.
The increased fixed cost reduces the entry of local suppliers and hence raises the prices of local intermediate goods.
Multinationals, even if they are not directly damaged, may find local sourcing unprofitable and leave the disaster-hit country, shrinking the local industry further.\footnote{If we allow for endogenous sourcing patterns of multinationals, those with heavily dependent on local suppliers may continue to stay in the disaster-hit country but reduce the intensity of local sourcing.
See Section 5.3.2 and Appendix 6.}
This mechanism helps understand the (sometimes long-lasting) negative disaster impact on FDI found in empirical studies (\citealp{EscalerasRegister2011}; \citealp{Doytch2020}; \citealp{TonerFriedt2020}; \citealp{Batalaetal2021}).
Natural disasters may also raise the cost of local sourcing through the destruction of transportation infrastructure or directly damage multinational plants.
In our model, the effects of these shocks are similar to those of the shock to the fixed cost for local suppliers because both reduce the profitability of multinationals staying in the host country.

Finally, conditions for multinationals showing resilience are identified.
Specifically, they are more likely to stay in the disaster-hit country, if local intermediate goods are more important in multinational production or trade costs of importing foreign inputs are lower.
Given that multinationals seeking low-cost inputs choose to enter the host country, a higher dependency on local suppliers and lower trade costs would lead to greater profits and thus prevent them from relocating.
Contrary to the warning by the Bank of Thailand, the 2011 Thailand floods did not cause long-lasting relocation and restructuring by manufacturing MNEs, perhaps because of their strong industrial linkages with local firms and progressive trade liberalization (\citealp{Milneretal2006}; \citealp{FelicianoDoytch2020}).\footnote{Using the information on Japan-Thailand bilateral input-output table, \cite{Milneretal2006} find a sizable and robust association between the number of Japanese affiliates in Thailand and their linkages with Thailand's industries.
\cite{FelicianoDoytch2020} document the recent development of Thailand's Free Trade Agreements.
Using Thailand's firm-level data, they find that reductions in import tariffs improve the performance of both local and foreign-owned firms.}

We further examine a number of extensions of the basic model including multinationals with heterogeneous productivity, gradual recovery from disasters, the role of the host country's market, endogenous sourcing patterns, and disaster risk.
They give similar results and additional implications.
Among others, we find that the least efficient multinationals are the first to leave the host country once a disaster hits.
Put differently, multinationals staying in the damaged country are likely to be the most efficient ones.
The heterogeneous-multinational setting avoids the extreme case where all multinationals completely exit.
The result may help understand the observation that in the aftermath of the 2011 floods, larger Japanese MNEs in Thailand did not change their local procurement share as much as smaller ones (\citealp{Hayakawaetal2015}).

\subsection{Relation to the literature}

We aim to contribute to the literature investigating the impact of multinationals on industrial development (\citealp{Alfaro2015} for a survey).
For example, using Irish micro-level data, \cite{GorgStrobl2002EER} find that the presence of multinationals promotes the entry of local manufacturing firms.\footnote{By contrast, studies reporting negative impacts include \cite{AitkenHarrison1999} for Venezuela; \cite{GorgStrobl2002EER} for Ireland; and \cite{Luetal2017} for China.}
Among many channels through which multinationals benefit host economies, an important one is vertical industrial linkages (e.g., \citealp{Javorcik2004}; \citealp{AlfaroUrenaetal2022}).
\cite{AlfaroUrenaetal2022} use firm-to-firm transaction data in Costa Rica and find that local firms increase their size and productivity after becoming a supplier to multinationals.

Although many theoretical attempts seek to understand the role of multinationals in vertically linked industries, only a limited number of studies have modeled (de)industrialization as switches between multiple equilibria (\citealp{RodriguezClare1996}; \citealp{MarkusenVenables1999}; \citealp{CarluccioFally2013}).\footnote{We differ from more recent theoretical studies on MNEs such as \cite{Alfaroetal2010}; \cite{Garetto2013}; \cite{RamondoRodriguez2013}; \cite{Arkolakisetal2018}; \cite{AdachiSaito2020}; and \cite{Gumpertetal2020} in that exogenous shocks always lead to a smooth change in equilibrium. In our model, by contrast, the shocks may bring a discontinuous jump.
Economic geography models deal with multiple equilibria but typically assume away distinction between local and multinational firms and/or input-output linkages (e.g., \citealp[Section 20.3.7]{ReddingTurner2015}; \citealp{Akamatsuetal2021}; \citealp{Gaspar2021}).
A few exceptions include \cite{FujitaThisse2006}; \cite{Hsuetal2020}; and \cite{KatoOkoshi2022}, although disaster impact is outside the scope of these studies.}
For example, \cite{MarkusenVenables1999} model the upstream and downstream industries and numerically illustrate the situation where the entry of downstream MNEs fosters local upstream industry.
Using a similar model, \cite{RodriguezClare1996} find that downstream MNEs benefit the host country if the intensity of their local sourcing is sufficiently high.
\cite{CarluccioFally2013} extend these two models further to consider a ``technological incompatibility" of local intermediates with multinational production.
These studies, however, analyze each equilibrium only separately and not explore which shocks trigger a shift from one equilibrium to another.

We rely on their framework and take one step further to analytically characterize the conditions under which an exogenous supply shock leads to an equilibrium switch and relate them to disaster impacts on multinationals.
Specifically, we simplify the model structure of \cite{MarkusenVenables1999} in a way such that the upstream and downstream industries are treated as one exhibiting roundabout production (\citealp{KrugmanVenables1995}) and explicitly introduce trade by MNEs.\footnote{The roundabout-production structure means that goods of one industry are used for production as intermediate goods in other industries including itself and are also consumed as final goods.
Notable applications include \cite{EatonKortum2002}; and \cite{CaliendoParro2015}.
The most advantage of modeling roundabout structure is to greatly simplify the analysis by not distinguishing between intermediate-good producers and final-good producers.
However, this comes at a cost of ignoring interesting aspects of inter-industry linkages such as the reallocation of resources between upstream and downstream sectors.}
In doing so, we can go beyond the numerical results of \cite{MarkusenVenables1999} and identify the conditions of MNEs showing resilience in terms of meaningful parameters such as the cost share of local intermediate goods and trade costs of foreign inputs.
Our simple model allows further extensions such as heterogeneous MNEs in productivity and gradual recovery from negative shocks, none of which is examined in the three studies.

Some studies examine the impact of various types of risk on multinational behavior (\citealp{Aizenman2003}; \citealp{AizenmanMarion2004}; \citealp{Russ2007}; \citealp{FillatGaretto2015}).
\cite{AizenmanMarion2004} investigate how the volatility of demand and supply shocks affect differently market-seeking FDI and efficiency-seeking FDI.
\cite{FillatGaretto2015} model multinational entry as a real-option problem, where multinationals show resilience against negative shocks because of the sunk cost they have paid to enter.
Unlike these studies focusing on risk under uncertainty, our model highlights actual physical damages, given the context of developing countries vulnerable to severe disasters (\citealp{ADB2013}; \citealp{SivapuramShaw2020}).
In an extended model with uncertainty,  we show that the risk of disasters alone may lead multinationals to leave the host country before a disaster actually occurs (see Section 5.3.3 and Appendix 7).

Simulation studies investigating disaster impact in an economy with industrial linkages are complementary to ours (\citealp{OkuyamaChang2004}; \citealp{Henrietetal2012}; \citealp{InoueTodo2019}; \citealp{GalbuseraGiannopoulos2018} for a survey).\footnote{For empirical studies on the impact of negative shocks including natural disasters in supply chains, see \cite{Todoetal2015}; \cite{BarrotSauvagnat2016}; \cite{Carvalhoetal2021}; \cite{Dhyneetal2021}; \cite{GigoutLondon2021}; and \cite{Kashiwagietal2021}.}
In an economy where buyer-supplier relationship constitutes a complex network, \cite{Henrietetal2012} show how network features such as concentration, clustering and connectedness between subregions either dampen or magnify the effects of a natural disaster.
We differ from these studies in terms of both in focus and modeling strategies.
They describe regional economies by fixing the location of production, whereas we emphasize the footloose-ness of internationally mobile MNEs in developing countries.
Although input-output linkages in our model are admittedly far simpler than theirs, we hope it serves as the first step toward a more comprehensive analysis on disasters and multinationals.  

This paper is also related to the vast body of empirical literature assessing the disaster impact on firm performance.\footnote{Empirical studies on the disaster impact on economic growth are also extensive.
For a summary of findings, see Tables 1 and 2 in \cite{CavalloNoy2011}; and Table 1 in \cite{FelbermayrGroschl2013}.
Theoretical contributions in this line include \cite{HallegatteDumas2009}; \cite{IkefujiHorii2012}; \cite{AkaoSakamoto2018}; and \cite{SchubertSmulders2019}.}
According to recent studies using micro-level data,  the destruction of physical capital due to natural disasters may be a good chance of upgrading it (\citealp{Leiteretal2009}; \citealp{VuNoy2018}; \citealp{Okazakietal2019}) or a bad one of decreasing productivity and the survival rates of firms (\citealp{Tanaka2015}; \citealp{Cainellietal2018}; \citealp{Coleetal2019}; \citealp{Meltzeretal2021}).\footnote{Whether surviving firms upgrade their capital may depend on their industry characteristics (\citealp{OkuboStrobl2021}).
Some studies report negative impacts on the export and import performance of affected firms at least in the short run (\citealp{AndoKimura2012}; \citealp{Boehmetal2019}; \citealp{Elliottetal2019}).}

Only a limited number of empirical studies examine the nexus between natural disasters and multinationals.
They find mixed results on whether the disaster effect is positive or negative and short- or long term (\citealp{EscalerasRegister2011}; \citealp{OhOetzel2011}; \citealp{Anuchitworawong2015}; \citealp{Doytch2020}; \citealp{TonerFriedt2020}; \citealp{Batalaetal2021}; \citealp{Neiseetal2021}).
\cite{Doytch2020} examines sectoral FDI inflows in 69 countries from 1980 to 2011.
Her dynamic panel regression results show negative effects in general, but also suggest that the effects vary depending on the type of disasters, industries, and regions.
Using data from foreign affiliates of 71 European MNEs, \cite{OhOetzel2011} find that major natural disasters have no significant impact on the number of affiliates, whereas terrorist attacks and technological disasters have negative impacts.
Closely examining disaster events in India in 2006--2019, \cite{TonerFriedt2020} report persistent intra-national shifts in multinationals' investment patterns from disaster-affected regions to non-affected regions.
Considering the discontinuous change in equilibrium due to a huge shock suggested by our theory, one may more likely find a negative effect of disasters with severe damages in sample countries less equipped with disaster prevention.

The remainder of the paper is structured as follows.
The next section presents a motivating empirical example on the disaster impact on FDI.
Section 3 describes the model and characterizes conditions under which different industrial configurations emerge.
Section 4 examines the impact of natural disasters and characterizes the conditions under which multinationals show resilience.
Section 5 provides two extensions of the basic model in detail and sketch three extensions briefly. 
The final section concludes the paper.

\

\section{Disaster impact on FDI: an empirical example}

To motivate theoretical analysis, this section provides suggestive evidence on natural disasters and FDI.
We first check whether natural disasters indeed affect FDI inflows.
We then confirm that the impact of disasters on FDI into developing countries is negative and persistent when their damages are severe.
We utilize country-level data on FDI in 1991--2015 from the World Bank Development Indicators (WDI) and data on natural disasters in 1976--2015 from the Emergency Disasters Database (EM-DAT) maintained by the Centre for Research on the Epidemiology of Disasters.\footnote{The EM-DAT database is commonly used in studies on natural disasters (e.g., \citealp{EscalerasRegister2011}; \citealp{FelbermayrGroschl2013, FelbermayrGroschl2014}; \citealp{Hallegatte2015}; \citealp{Doytch2020};  \citealp{KikkawaSasahara2020})}
Detailed information on data sources and summary statistics are provided in Data Appendix.

We regress FDI inflows from the world to country $i$ in year $t$ on the total number of natural-disaster events that hit country $i$ in $t-1$ to $t-\tau$ years.\footnote{The specification is similar to the one by \cite{EscalerasRegister2011}.
The key differences, however, are that (a)we use FDI inflows rather than FDI inflows divided by GDP; that (b)we include dummies for developing country; and that (c)the accumulated number of prior disasters is decomposed into recent and past ones in Table 2.}
Because the disaster impact on FDI may differ depending on its intensity, we distinguish between severe and non-severe disasters.
A disaster event is defined as a severe one if it records financial damage as a share of GDP exceeding its median value for all countries that have ever suffered from disasters in 1981--2015.
Hence, a non-severe event is defined as one if it is equal to or below the median.
In addition, to see whether disaster impacts differ in the level of development, we interact the accumulated number of disasters with the dummy for developing countries.
The regression equation is 
\begin{align*}
\text{FDI}_{i, t} = \alpha &+ \beta \sum_{s=1}^{\tau} (\text{No. of natural disasters})_{i, t-s} 
\\
&+ \gamma \sum_{s=1}^{\tau} (\text{No. of natural disasters})_{i, t-s} \times (\text{Developing})_{i, t} \\
\
&+ (\text{Developing})_{i, t} + \bm{x}'_{i,t} \bm{\delta} + \eta_i + \theta_t +u_{i, t}, %\tag{1}
\end{align*}
where $\tau$ takes 1, 5, or 10 years; $(\text{Developing})_{i, t}$ is the dummy for developing country;\footnote{We define developed and developing countries according to the classification of the World Bank; developed countries are those belonging to the high-income group, while developing countries to the other income groups.} $\eta_t$ is the country fixed effect; $\theta_t$ is the year fixed effect; and $u_{i, t}$ is the error term.
The column vector $\bm{x}_{i, t}$ includes control variables of the host country, which are considered as essential determinants of FDI in both the theoretical and empirical literature (\citealp{Markusen2004}; \citealp{AntrasYeaple2014};  \citealp{BlonigenPiger2014} for reviews), i.e., the log of real GDP (a measure of market size), the average tariff rate of manufacturing products (a measure of the inverse of trade openness), and the unit labor cost (a measure of labor costs).
As FDI seeks greater demand and lower priced factors, we expect that GDP has a positive effect and the unit labor cost has a negative effect on FDI inflows.
The openness to trade has a mixed effect and may increase or decrease FDI inflows.\footnote{On the one hand, low trade openness may increase the entry of  multinationals who try to be more proximate to consumers in the host market.
On the other hand, it may decrease the entry of multinationals who try to export from the host country using low cost factors there.
The former case captures motivations for market seeking, or ``horizontal'' FDI, while the latter case for efficiency seeking, or ``vertical'' FDI.
The effect of trade openness on FDI may become more complex when considering other types of FDI.
See the suggested references in the text for details.}
We use one-year lag of these control variables.

Regression results are shown in Table 1.\footnote{For completeness, we provide the results of regressions without developing-country dummies in Tables A4 and A5 in Data Appendix.}
In columns (1) to (3), the disaster variables  count all events regardless of their intensity, while in columns (4) to (6) they count only severe ones.
In the first three columns of Table 1, the coefficients on the disaster variables are all positive and significant, i.e., $\widehat{\beta}>0$.
This may reflect the facts that FDI temporarily away from the disaster-hit economy comes back as it recovers or that natural disasters provide an opportunity for foreign investors to upgrade capital equipment (see \citealp{Doytch2020}; and \citealp{Neiseetal2021} for similar findings).
The interaction of the disaster variable with the developing-country dummy has a negative and significant coefficient in column (1), i.e., $\widehat{\gamma}<0$.
The positive effect of prior disasters on current FDI could be smaller for developing countries than for developed countries, although their base level of FDI is higher, as indicated by the positive and significant coefficients on the developing-country dummy in all columns.

Such a heterogeneous disaster impact depending on the level of economic development is more articulated when we focus on severe events, as shown in columns (4) to (6) of Table 1.
The level term of the disaster variable has an  insignificant coefficient in the three columns, while its interaction with the developing-country dummy has a negative and significant coefficient in columns (5) and (6), i.e., $\widehat{\gamma}<0$.
FDI into developing countries negatively respond to severe disasters in the last 5 or 10 years, while FDI into developed countries do not.
This suggests the vulnerability of developing countries to disasters possibly due to insufficient resources for disaster prevention. 

In all columns of Table 1, the lagged value of $\log (GDP)$ is consistently positive and significant, as we expected.\footnote{Although the variables of our primary interest are the disaster ones, one may be concerned about the endogenity of GDP.
The results of regressions without GDP are quite similar to those in Tables 1 and 2.}
The lagged value of \textit{Tariff}, a measure of the inverse of trade openness, is consistently negative and largely significant.
This may suggest FDI in our sample uses their host country as a base for exporting to the other countries.
The lagged value of $\textit{Unit labor cost}$ is not significant and unexpectedly positive in some cases.

\pagebreak

\begingroup
\begin{center}
Table 1. \ Disaster impact on FDI
\includegraphics[scale=0.8]{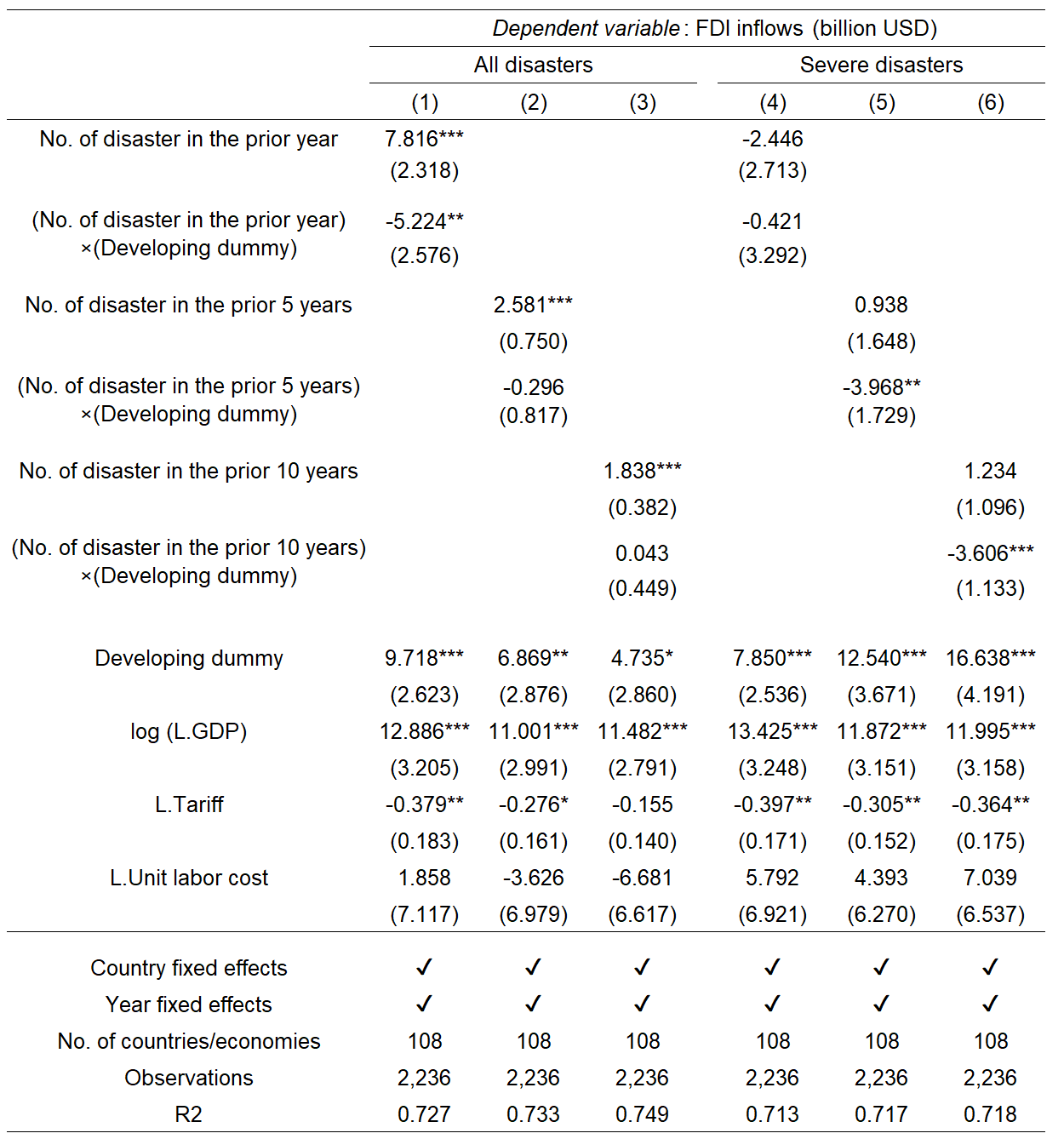}
\end{center}
\vspace{-0.3cm}
\begin{spacing}{1.0}
\noindent {\small \textit{Notes:} Robust standard errors clustered at region$\times$year ($5 \times 25$) are in parentheses.
The sample period is from 1991 to 2015.
FDI inflows (billion USD) are deflated by the price level of real GDP.
Developed countries are those belonging to the high-income group, while developing countries to the other income groups, according to the classification of the World Bank.
A severe disaster in country $i$ in year $t$ is defined as one that records financial damages as a share of GDP exceeding its median for all countries that have ever experienced natural disasters in 1981--2015.
A non-severe disaster in country $i$ in year $t$ is defined as one that is not severe.
The three control variables, the log of real GDP, the average tariff rate, and the unit labor cost, are lagged one year.
We exclude tax haven countries identified by \cite{HinesRice1994}.
\\
$^{*}$Significant at 10$\%$ level; $^{**}$Significant at 5$\%$ level; $^{***}$Significant at 1$\%$ level. }
\end{spacing}
\endgroup

\

We then explore potential long-lasting effects of disasters by decomposing the number of disaster events in the prior 10 years into recent and past ones.
Namely, the modified estimation equation includes three disaster variables: (i)the number of events in the prior year, (ii)that in the prior 2 to 5 years, and (iii)that in the prior 6 to 10 years.
We also take into account heterogeneous disaster effects depending on the level of development and the intensity of disasters.
Table 2 shows the regression results, where the disaster variables in column (1) count all events and those in column (2) count only severe events.
Focusing on statistically significant coefficients in column (1), we see that FDI inflows to developing countries negatively respond to all types of disasters in the prior year ($5.943-6.182=-0.239$).
In column (2), by contrast, they negatively respond to severe disasters in the prior 2 to 5 years ($2.043-5.094=-3.051$) and weakly do so to those in the prior 6 to 10 years ($1.387-3.079=-1.692$).
These results suggest that if developing countries are hit by large scale disasters,  their negative impact on FDI may continue in the medium and possibly long run.

These observations guide our theory in a way such that the negative long-term impact of disasters could be modeled as a switch between multiple equilibria of industrial configurations.
Important aspects are that the equilibrium switch may be likely to occur in developing counties and to be brought especially by severe disasters.
Building a model capturing these features, our aim is then to characterize the conditions under which FDI in developing countries shows resilience against substantial disasters.\footnote{In the basic model, the disaster impact is modeled as the exit of multinationals from the host country.
In the regression analysis, we cannot tell withdrawn investments from our country-level FDI data.
The negative coefficients on the disaster variables mean both that disasters deter new investments and that they force existing investments to exit.
Ideally, we would use affiliate-level data and see the effect of disasters on the exit probability of existing affiliates.
Instead, we provide an extended model consistent with the regression analysis in Section 5.2 and Appendix 4.
It models new investment flows as the reentry of multinationals into the disaster-hit country.}

\pagebreak

\begin{center}
Table 2. \ Disaster impact on FDI: recent vs. past disasters
\includegraphics[scale=0.75]{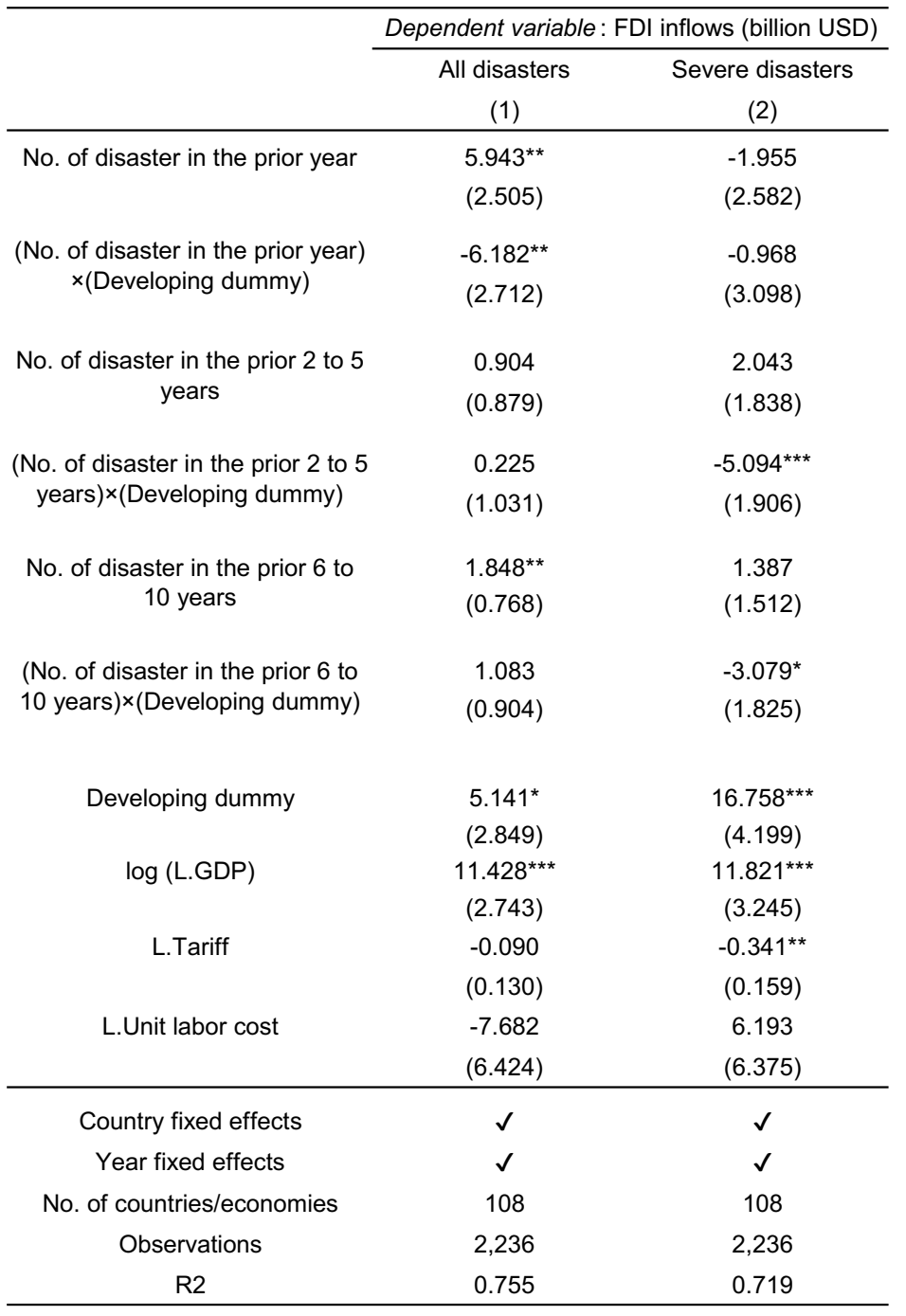}
\end{center}
\vspace{-0.3cm}
\begin{spacing}{1.0}
\noindent {\small \textit{Notes:} Robust standard errors clustered at region$\times$year ($5 \times 25$) are in parentheses.
The sample period is from 1991 to 2015.
FDI inflows (billion USD) are deflated by the price level of real GDP.
Developed countries are those belonging to the high-income group, while developing countries to the other income groups, according to the classification of the World Bank.
A severe disaster in country $i$ in year $t$ is defined as one that records financial damages as a share of GDP exceeding its median for all countries that have ever experienced natural disasters in 1981--2015.
A non-severe disaster in country $i$ in year $t$ is defined as one that is not severe.
The three control variables, the log of real GDP, the average tariff rate, and the unit labor cost, are lagged one year.
We exclude tax haven countries identified by \cite{HinesRice1994}.
\\
$^{*}$Significant at 10$\%$ level; $^{**}$Significant at 5$\%$ level; $^{***}$Significant at 1$\%$ level. }
\end{spacing}

\

\

\section{The model}

Consider the host country and the foreign country, which comprises the rest of the world.
There are two sectors: a differentiated-product sector and a homogeneous-product sector.
In the differentiated sector, there are three types of firms: host domestic firms, multinationals, and foreign domestic firms.
The last two are foreign-owned firms.
Differentiated varieties have two roles: final goods for consumers in the two countries and intermediate goods for host domestic and multinational firms.
To best fit our model in the context of developing countries, the host-country market is assumed to be small and localized and thus served by only host domestic firms, while the foreign-owned firms serve the foreign country's market.\footnote{We allow multinationals to serve both markets in the host and foreign countries in Section 5.3.1 and Appendix 5.}
Foreign capital used as fixed costs of setting up foreign-owned firms is free to choose between the host and the foreign countries.
The structure of our model, depicted in Fig. 1, is close to those of \cite{MarkusenVenables1999}; and \cite{CarluccioFally2013}.
The crucial difference is, however, that unlike their models we combine local upstream and downstream firms into one local producer using roundabout production technology (\citealp{KrugmanVenables1995}).\footnote{Another important difference of our model from theirs is that we introduce multinationals' imports of foreign intermediates, as we will see below.}
This simplification greatly helps us obtain analytical results, while maintaining a similar mechanism.

In the homogeneous sector, firms use labor to produce a homogeneous good.
We assume that the good is freely traded and the unit labor requirement is unity.
These imply that the good's price is equal to both a constant world price and wage.
We choose the homogeneous good as the num{\'e}raire, so that both the good price and the wage are unity.
Labor is freely mobile between two sectors, which equalizes wages between the sectors.

\

\begin{center}
\includegraphics[scale=1.1]{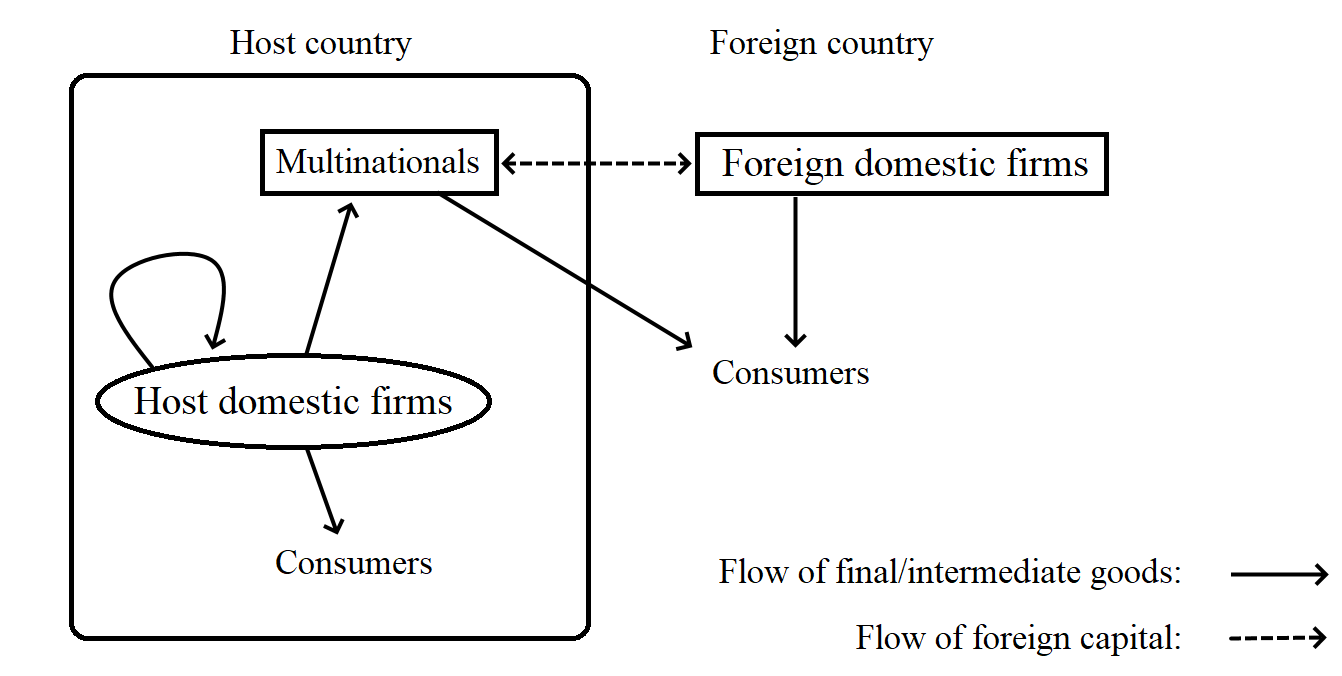} \\
Fig. 1. \ Model structure
\end{center}

\

\subsection{Tastes and production}

\textit{Consumers.} \ \ The representative consumer in the host country has the following
utility function: 
\begin{align*}
&u=Q^{\alpha} (q^O)^{1-\alpha}, \tag{1} \\ 
\
&\text{where} \ \ Q=\left( \int q^c(\omega )^{\frac{\sigma -1}{\sigma }}d\omega \right)^{\frac{\sigma }{\sigma -1}},
\end{align*}
and where $Q$ is the composite of the differentiated final goods and $q^c(\omega)$ is the consumer demand for an individual variety of $\omega$ produced by the host domestic firms.
$q^O$ is the demand for the num{\'e}raire good, 
$\alpha$ is the expenditure share on final goods,
and $\sigma>1$ measures the elasticity of substitution between varieties. 
We solve the first-order condition (FOC) of utility maximization to obtain demand function for variety $\omega$:
\begin{align*}
&q^c(\omega)=\left( \frac{p(\omega )}{P}\right) ^{-\sigma}\frac{E}{P}, \tag{2} \\
\
&\text{where}\ \ P=\left( \int p(\omega )^{ 1-\sigma}d\omega \right)^{\frac{1 }{1-\sigma}},
\end{align*}
and where $p$ is the price of the variety and $E$ is the aggregate expenditure on differentiated goods, which is equal to a $\alpha$ share of income.
The income of the representative consumer consists of labor income and excess profits repatriated by host domestic firms (if not zero).
Noting that host domestic firms are symmetric, the price index, $P$, becomes
\begin{align*}
&P=\left(N p^{1-\sigma }\right) ^{\frac{1}{1-\sigma }} = N ^{\frac{1}{1-\sigma }}  p, \tag{3}
\end{align*}
where $N$ is the number (or mass) of host domestic firms.
We will suppress the variety index of $\omega$ in what follows.

Our focus is on the host country; hence, we do not describe in detail the market in the foreign country.
The demand functions for a typical variety produced by multinational firms, $i=m$, and foreign domestic firms, $i=f$, are 
\begin{align*}
&q_i = \left( \frac{p_i}{P^*} \right)^{-\sigma} \frac{E^*}{P^*} \equiv p_i^{-\sigma} D^*, \ \ \ \ i \in \{ m, f\}, \tag{4}
\end{align*}
where $P^*$ and $E^*$ are respectively the price index and the total expenditure on differentiated goods in the foreign country.
The market size of the foreign country is summarized by a constant $D^* \equiv E^* (P^*)^{\sigma-1}$.

\textit{Host domestic firms.} \ \ Each host domestic firm requires $F$ amounts of labor for setup costs.
Once established, they use $\widetilde{a}$ units of a Cobb-Douglas composite input to produce one unit of output.
The composite input comprises foreign intermediate goods (with a share $1-\mu$) and local intermediate goods produced by the differentiated sector itself (with a share $\mu$).
The price of foreign intermediates is exogenously given by $\tau p_u^*$, where $p_u^* \ge 1$ is a constant free-on-board price and $\tau \ge 1$ represents trade costs between the host country and the foreign country.\footnote{By contrast, we assume that trade in final goods is freely shipped.
Introducing trade costs in final goods would not change the qualitative results because trade costs both in final and intermediate goods work in the same way that raises the selling price $p_m$ of multinationals.}
The minimized total cost to produce $q$ units of a typical variety is then
\begin{align*}
&C(q)=P^{\mu} w^{1-\mu} \widetilde{a} q + wF = P^{\mu} \widetilde{a} q + F, \tag{5}
\end{align*}
noting that the wage is equal to one: $w=1$.
As the price elasticity of demand is given by $\sigma$, which we will see later, the FOCs for profit maximization yield the usual constant-markup pricing: 
\begin{align*}
p = \frac{\sigma \widetilde{a} P^{\mu} }{\sigma-1} = a P^{\mu}, \tag{6}
\end{align*}
where we choose $\widetilde{a}$ such that $\widetilde{a} = (\sigma-1)a/\sigma$.
From Eqs. (3) and (6), we can rewrite the price index as
\begin{align*}
P =a^{\frac{1}{1-\mu}} N^{\frac{1}{(1-\sigma)(1-\mu)}}, \tag{7}
\end{align*}
which decreases with $N$ because $\sigma>1$ and $\mu \in (0, 1)$,

\textit{Multinationals.} \ \ Each multinational firm needs one unit of foreign capital to start operation and $\widetilde{a}_m$ units of a composite input per unit of output.
The composite input for multinationals is different from that for host domestic firms and is made up of foreign-produced intermediate goods (with a share $1-\mu_m$) and locally-produced intermediate goods (with a share $\mu_m$).
The cost function for a typical multinational is then
\begin{align*}
&C_{m}(q_{m})= P^{\mu_m} (\tau p_u^*)^{1-\mu_m} \widetilde{a}_m q_m +r_{m}, \tag{8}
\end{align*}
where $q_m$ is given by Eq. (4) and $r_{m}$ is the rental rate of capital.
The FOCs give the optimal price:
\begin{align*}
p_m = \frac{\sigma \widetilde{a}_m P^{\mu_m} (\tau p_u^*)^{1-\mu_m}}{\sigma-1} = a_m P^{\mu_m} (\tau p_u^*)^{1-\mu_m}, \tag{9}
\end{align*}
where $\widetilde{a}_m$ is chosen such that $\widetilde{a}_m = (\sigma-1)a_m/\sigma$.

\textit{Foreign domestic firms.} \ \ Each foreign domestic firm needs one unit of foreign capital for setup costs and $\widetilde{a}_m$ units of foreign intermediate goods per unit of output.
They source all inputs from the foreign country's suppliers, who we do not explicitly model, without incurring trade costs.
The cost function for a typical foreign domestic firm is
\begin{align*}
C_f(q_f)=p_u^* \widetilde{a}_m q_f+r_f, \tag{10}
\end{align*}%
where $r_f$ is the rental rate of foreign capital for foreign firms.
The optimal price for the variety is simply given by $p_f = a_m p_u^*$.

Finally, we need to consider clearing conditions for goods and labor markets.
The total output for the variety produced by each host domestic firm, $q$, must be equal to the sum of consumer demand and the intermediate-input demand by multinationals and host domestic firms themselves.
By using Eq. (2) and applying Shepard's lemma to Eqs. (5) and (7), we obtain
\begin{align*}
q &= \left( \frac{p}{P} \right)^{-\sigma} \frac{E}{P} + N \frac{\partial C(q)}{\partial p} + N_m \frac{\partial C_m(q_m)}{\partial p} \\
\
&= p^{-\sigma} \left[ P^{\sigma-1}E  + N \mu P^{\sigma +\mu -1} \widetilde{a} q + N_m \mu_m P^{\sigma +\mu_m -1} (\tau p_u^*)^{1-\mu_m} \widetilde{a}_m q_m \right], \tag{11}
\end{align*}
noting that the demand elasticity is equal to the elasticity of substitution between varieties $\sigma$.

Similarly, the labor demand in the differentiated sector can be derived by applying Shepard's lemma to Eq. (5): $N \cdot \partial C(q)/\partial w = N \left[ (1-\mu) P^{\mu} \widetilde{a}q + F \right]$.
The labor-market clearing requires that the sum of the labor demand in both the differentiated and the num{\'e}raire sectors must be equal to the total workforce in the host country, $L$.
It determines the upper bound of $N$ such that $N \le \overline{N} \equiv L/F[\sigma(1-\mu)+\mu]$ (see Appendix 1).
In the num{\'e}raire sector, its labor demand and thus its outputs are adjusted in a way such that imports/exports of the good are balanced.

\subsection{Industry equilibrium}

All three types of firms freely enter and exit from the differentiated sector.
Let $\Pi \equiv pq -C(q)$ be the excess profit of domestic firms, i.e., operating profits minus fixed costs.
We assume a gradual entry-and-exit process governed by the law of motion such that $\dot{N} = \Pi$, where the dot denotes the time derivative.\footnote{More rigorous dynamic analyses of location models can be found in \cite{Baldwin2001}; \cite{Ottaviano2001}; \cite{Boucekkineetal2013}; and \cite{FujishimaOyama2021}.}
Entry occurs when $\Pi>0$, while exit occurs when $\Pi<0$.
This adjustment process stops at the point where host domestic firms break even, that is, $\Pi = 0$.
Using Eqs. (5), (6), and (10), we can derive the combinations of $N$ and $N_m$ that satisfy $\Pi = 0$ as
\begin{align*}
&\Pi = pq -C(q) = pq - P^{\mu} \widetilde{a} q - F = 0, \\
\
&\to N_m = \Theta N^{-\frac{\mu_m}{1-\mu}}  \left( N  -\alpha \overline{N} \right), \tag{12} \\
\
&\text{where} \ \ \Theta \equiv \frac{ \sigma F  [\sigma(1-\mu)+\mu] \left[ a_m a^{\frac{\mu_m}{1-\mu}}  (\tau p_u^*)^{1-\mu_m} \right]^{\sigma-1} }{\mu_m D^*(\sigma-1) }, \ \ \ \
\overline{N} \equiv \frac{L}{F[\sigma(1-\mu) +\sigma]},
\end{align*}
noting that $\overline{N}$ represents the maximum number of host domestic firms, which depends on its workforce, $L$.

We draw a typical $\Pi=0$-locus in Fig. 2(a) with arrows indicating the direction of motions.\footnote{At $N=\overline{N}$, the host domestic firms use up all labor.
Since no further entry is possible, the incumbent firms make positive profits.}
The locus has an upward slope under a reasonable condition that the expenditure share on varieties is not sufficiently small such  that $\alpha > 1 - (1-\mu)/\mu_m$, which we will assume throughout the paper.\footnote{The $\Pi=0$-locus increases with $N$ if $\partial N_m/\partial N = \Theta N^{-1-\frac{\mu_m}{1-\mu}}[\mu_m \alpha \overline{N} +(1-\mu-\mu_m)N]/(1-\mu)>0$, or equivalently $\mu_m \alpha \overline{N} +(1-\mu-\mu_m)N>0$ holds.
A sufficient condition for this inequality is $\alpha > 1 - (1-\mu)/\mu_m$.} 
The reason for the upward sloping locus is as follows.
Suppose that the host economy is initially at the point on the $\Pi=0$-locus and then experiences an increase in the number of multinationals.
This multinational entry generates positive profits by raising the intermediate-good demand.
To maintain zero profits, a greater number of host domestic firms is necessary.

We turn to multinationals and foreign domestic firms.
Let $\Pi_i = p_i q_i - C_i(q_i)$ be the excess profit of firm $i \in \{ m, f\}$.
Then, free entry and exit lead to $\Pi_i=0$, where the rental rate of foreign capital is exactly covered by operating profits: $r_i = p_i q_i/\sigma$.
As a result of the arbitrage behavior of foreign investors, foreign capital chooses the type of firms that generates higher rental rate/operating profits.
Foreign capital becomes indifferent between becoming a multinational and a foreign domestic firm when the capital-return differential is zero:
\begin{align*}
&\Delta r_m \equiv r_m - r_f = p_m q_m/\sigma - p_f q_f/\sigma = 0, \\
\
&\to N =  a^{\sigma-1} \left[ \tau^{\frac{1-\mu_m}{\mu_m}} (p_u^*)^{-1} \right]^{(\sigma-1)(1-\mu)} \equiv N_0, \tag{13}
\end{align*}
where we used Eqs. (4), (8) to (10) and $p_f = a_m p_u^*$.
A typical $\Delta r_m = 0$-locus is drawn in Fig. 2(b) with arrows indicating the direction of motion, where $K_f$ is the total amount of foreign capital.
The locus does not depend on the number of multinationals, $N_m$, because the operating profits of both types of firms are independent of $N_m$.
What matters for the location decision of foreign capital is the price index of local intermediate goods, $P$, which depends only on $N$.
The $\Delta r_m = 0$-locus, or equivalently the $N=N_0$-line is the threshold number of local suppliers above which foreign capital chooses to enter the host country.

As in the entry-and-exit of host domestic firms, we assume the gradual relocation process of foreign capital such that $\dot{N}_m=\Delta r_m$.
When $\Delta r_m>0$ or $N > N_0$, multinationals can produce at a lower marginal cost than foreign domestic firms, and thus all foreign capital eventually chooses to become the multinational, i.e., $N_m=K_f$.
When $\Delta r_m<0$ or $N<N_0$, the opposite is true and no foreign capital enters the host country, i.e., $N_m=0$.

\

\begin{center}
\includegraphics[scale=0.8]{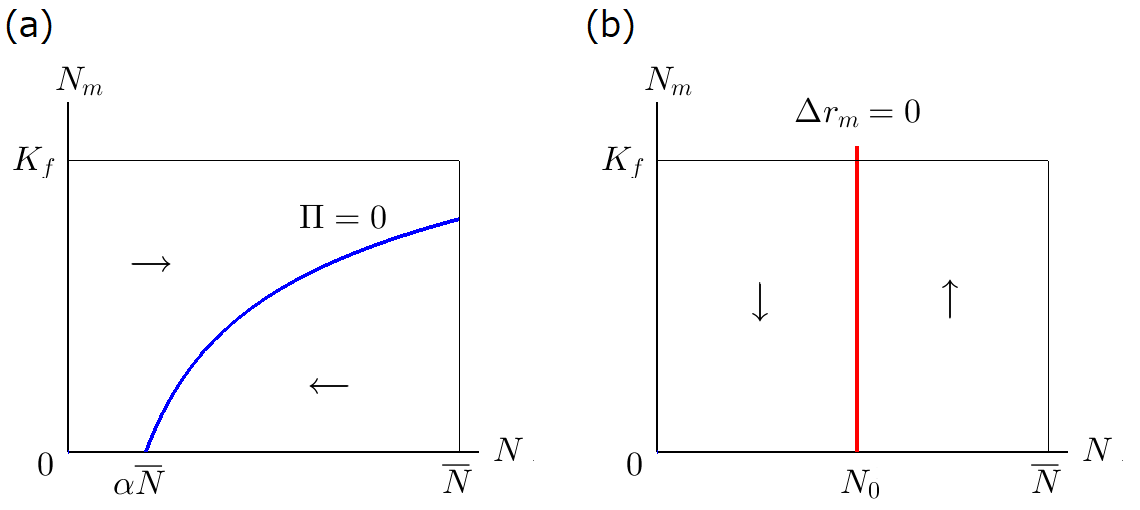} \\
Fig. 2. \ Equilibrium curves
\end{center}

\

Industry equilibrium, in which both the adjustments of $N$ and $N_m$ are completed, is determined by the relative positions of the two curves.
When $N_0 < \alpha \overline{N}$, foreign capital finds it more profitable to become the multinational than to become the foreign domestic firm even if there are few host domestic firms and hence $P$ is high.
More multinationals demand local intermediates, inducing more local suppliers to enter.
Consequently, the host economy reaches point $S_1: (N, N_m)=(\overline{N}, K_f)$, a stable equilibrium where both multinationals and host domestic firms coexist.

Meanwhile, when $N_0 > \overline{N}$, no foreign capital prefers to become the multinational.
The unique stable equilibrium is point $S_2: (N, N_m)=(\alpha \overline{N}, 0)$, where only local firms operate in the host country.
Since there are no intermediate-input demand by multinationals, the number of host domestic firms, $\alpha \overline{N} = \alpha L/F[\sigma(1-\mu)+\sigma]$, is constrained solely by the local expenditure on differentiated goods, $E=\alpha L$.

When $N_0$ is in between $\alpha \overline{N}$ and $\overline{N}$, both points, $S_1$ and $S_2$, are stable equilibria.
Fig. 3 shows this situation.
The foreign-capital-return differential is not large enough; hence, whether to enter the host country or to remain in the foreign country has no definite answer.
Rather, the answer depends on the number of local firms currently prevailing in the host country.
This complementarity between multinationals and host domestic firms creates a coordination problem.
Foreign capital will enter the host country if local production is expected to expand.
If there exists a prospect of multinationals increasing intermediate-input demand, more local suppliers will enter, leading the host economy to point $S_1$.
The same reasoning applies to point $S_2$: local intermediate production is never expected to
expand, if there is no prospect of multinational entry.
Which of the two equilibria arises depends on the expectations of firms/investors and the initial industry configuration (\citealp{Krugman1991QJE}; \citealp{Matsuyama1991}).  
Point $U$ is a saddle but an unstable equilibrium, because our model does not include any jump variables to put the host economy on the saddle path.

\

\begin{center}
\includegraphics[scale=0.8]{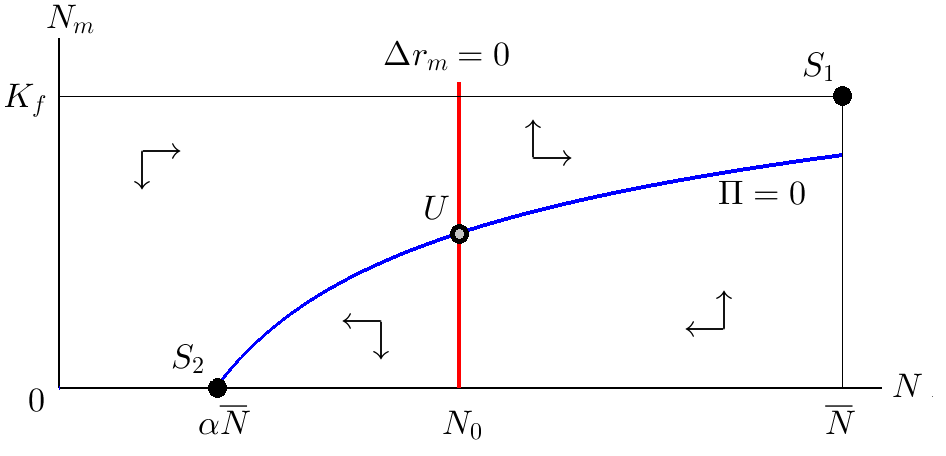} \\
Fig. 3. \ Multiple equilibria
\end{center}

\

As $\overline{N}$ is a function of fixed labor input $F$, it is convenient to characterize industry equilibrium using $F$.
We assume that (a) the expenditure share on varieties is not too small such that $\alpha>1-(1-\mu)/\mu_m$, in which case the $\Pi=0$-locus has an upward slope, and (b) it never touches the upper limit of $N_m$, that is, a sufficiently large amount of foreign capital such that $K_f > \Theta \overline{N}^{\frac{1-\mu-\mu_m}{1-\mu}} (1-\alpha)$ (see Eq. (12)).
Then we obtain the following proposition.

\

\noindent
{\bf Proposition 1 (Industry equilibrium).} \ \ {\it Using the fixed labor requirement for host domestic firms, $F$, industry equilibrium is characterized as follows:
\begin{itemize}
\item[(i)] If $F$ is small such that $F < F_a \equiv \alpha L/N_0[ \sigma(1-\mu)+\mu]$, the configuration in which multinationals and host domestic firms coexist, $S_1: (N, N_m)=(\overline{N}, K_f)$, is the only stable equilibrium.
\item[(ii)] If $F$ is large such that $F > F_b \equiv L/N_0[ \sigma(1-\mu)+\mu]$, the configuration in which all multinationals leave, $S_2: (N, N_m)=(\alpha \overline{N}, 0)$, is the only stable equilibrium.
\item[(iii)] If $F$ is intermediate such that $F_a \le F \le F_b$, both configurations are stable equilibria.
\end{itemize}
}

\

\section{Natural disasters and the resilience of FDI}

\subsection{Adverse shock to local firms}

Let us now consider natural disasters in the host country.
A disaster impact is modeled as an adverse shock to the fixed labor input for host domestic firms $F$, that is, an increase from $F$ to $F + \Delta F$ with $\Delta F>0$.
We examine (i) whether the natural disaster changes the equilibrium configuration, and if so (ii) when such a change is less likely to occur: that is, when the host economy is more likely to be resilient. 

Suppose $F$ is in between $F_a$ and $F_b$ and that the host economy is at point $S_1: (N, N_m)=(\overline{N}, K_f)$.
Fig. 4(a) depicts this situation and marks the initial point with a double circle.
As shown by the thin dashed line in Fig. 4(b), an increase in $F$ raises labor used in each host domestic firm and hence reduces the maximum number of local suppliers the host country can support, i.e., a decrease from $\overline{N}$ to $\overline{N}'$.
It means that more host domestic firms become unprofitable, moving also the $\Pi=0$-locus left.
The $\Delta r_m=0$-locus does not change, however, as $F$ does not directly enter the operating profits of multinationals and foreign domestic firms.

In Fig. 4(b), the shock is so substantial that the new $N=\overline{N}'$-line is located to the left of the $\Delta r_m=0$-locus.
The decline in local supplying industry raises input prices and thus makes multinationals unprofitable.
If such an adverse effect reaches a certain threshold, multinationals suddenly start leaving the host country.
Multinational exits in turn decreases demand for local intermediates and causes a further decline in the local industry.
The dotted arrow in Fig. 4(b) traces the industry evolution: it goes left horizontally up to $N_0$ and then goes lower left, heading for a new equilibrium at point $S_2': (N, N_m)=(\alpha \overline{N}', 0)$, as indicated by a double circle in Fig. 4(c).

\

\begin{center}
\includegraphics[scale=1.5]{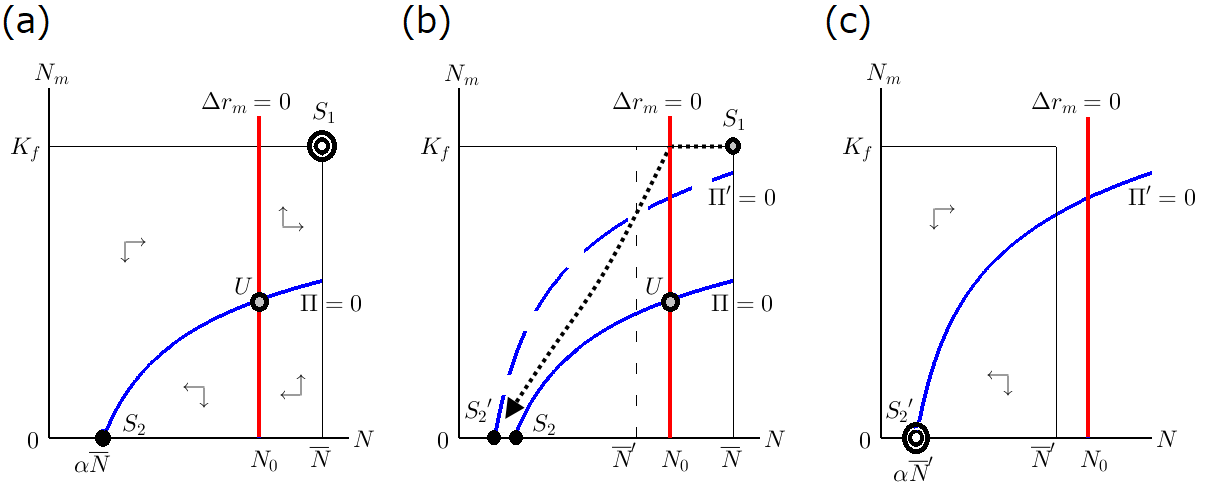} \\
Fig. 4. \ Equilibrium switch due to a natural disaster (from panel (a) to (c))
\end{center}

\

The condition for the equilibrium switch to occur is
\begin{align*}
&\overline{N}' \equiv \frac{L}{(F +\Delta F)[\sigma(1-\mu) + \mu]} < N_0 \equiv a^{\sigma-1} \left[ \tau^{\frac{1-\mu_m}{\mu_m}} (p_u^*)^{-1} \right]^{(\sigma-1)(1-\mu)}, \\
\
&\to F +\Delta F > F_b \equiv  \frac{L}{N_0[\sigma(1-\mu) + \mu]}, \\
\
&\to \Delta F > F_b - F \equiv \Delta F_{min}. \tag{14}
\end{align*}
A natural disaster causes a shift in equilibrium only if the magnitude of shock exceeds a certain threshold.
In addition, as long as the location decision is myopic, the equilibrium shift is permanent.\footnote{We consider a forward-looking location decision in Sections 5.2 and 5.3.3.}
This result echoes the long-lasting negative impact of severe disasters on FDI found by \cite{TonerFriedt2020} and the motivational observations we laid out in Table 2.

We can see that the threshold level of shock, $\Delta F_{min}$, increases with the cost share of local intermediate goods in multinational production $\mu_m$, and decreases with trade costs $\tau$.
Put differently, as $\mu_m$ is higher or $\tau$ is lower, the equilibrium switch is less likely to occur.
Multinationals more dependent on low-price local intermediates (higher $\mu_m$) make greater profits, so that they are more likely to stay in the host country despite the decreased number of local suppliers.\footnote{Under the situation where point $S_1$ can be an equilibrium, the sourcing costs of local intermediates are not higher than those of foreign intermediates, i.e., $P \le \tau p_u^*$ (see Appendix 2 for the proof).
In contrast to $\mu_m$, the host domestic firm's cost share of local intermediate goods, $\mu$, has an ambiguous effect on $\Delta F_{min}$.
A higher $\mu$ means that host domestic firms use more composite intermediate goods and use less labor.
More labor can be devoted to the fixed labor requirement for potential local firms, leading to more entry and thus making the price index of local intermediates lower.
However, a higher $\mu$ may push the price index up if the local intermediates are more expensive than labor.
If the productivity of host domestic firms are sufficiently high such that $a \le e^{\frac{1}{\sigma}} (\alpha \overline{N})^{\frac{1}{\sigma-1}}$, where $e$ is the base of natural log, the former effect dominates the latter, so that $\partial (\Delta F_{min})/\partial \mu > 0$ holds.
See Appendix 2 for details.}
Lower trade costs play a similar role by reducing the import price of foreign intermediates and thus making multinationals more profitable.
Contrary to the concern of the Bank of Thailand, the 2011 flood did not cause drastic long-run changes in the location and production of manufacturing MNEs (\citealp{HaraguchiLall2015}).
This may reflect the fact that Thailand had already established strong linkages with multinationals and engaged in trade liberalization (\citealp{Milneretal2006}; \citealp{FelicianoDoytch2020}).

Assuming (a) $\alpha>1-(1-\mu)/\mu_m$ and (b) $K_f > \Theta \overline{N}^{\frac{1-\mu-\mu_m}{1-\mu}} (1-\alpha)$ as in Proposition 1, we can prove the following proposition (see Appendix 2 for the proof).

\

\noindent
{\bf Proposition 2 (Natural disaster).} \ \ {\it Suppose that the fixed labor input, $F$, is intermediate so that multiple equilibria arise, i.e., $F \in [F_a, F_b]$, and that the host economy is initially at point $S_1$, where multinationals and host domestic firms coexist.
If $F$ increases to $F+\Delta F$ due to a natural disaster, then the following holds:
\begin{itemize}
\item[(i)] If the level of shock is substantial such that $F+\Delta F > F_b$, or equivalently $\Delta F > \Delta F_{min}$, the equilibrium switches from $S_1$ to $S_2'$, where the host country loses all multinationals.
\item[(ii)] The threshold level of shock increases with the cost share of local intermediate goods in multinational production ($\mu_m$) and decreases with trade costs ($\tau$), i.e., $\partial (\Delta F_{min})/\partial \mu_m \ge 0$ and $\partial (\Delta F_{min})/\partial \tau \le 0$, where equality holds at zero trade costs ($\tau=1$).
That is, the natural disaster is less likely to trigger the equilibrium switch if multinationals are more dependent on local intermediate goods or they face lower trade costs.
\end{itemize}
}

\

\subsection{Other types of shock}

The essence of the above analysis is that the negative shock to local suppliers raises the cost of local sourcing and hence discourages foreign capital to stay in the host country.
Other types of shocks that directly or indirectly raise the local-input price would give qualitatively the same result. 
For example, one can think of additional intra-national trade costs that local suppliers have to incur when delivering varieties to multinationals and consumers.
Through the destruction of domestic transportation infrastructure, natural disasters may raise intra-national trade costs, leading to a higher local-input price.
This translates into an upward shift of the $\Pi=0$-locus and a rightward shift of the $\Delta r_m=0$-locus (i.e., the $N=N_0$-line); however, it does not change the $N=\overline{N}$-line unlike the case in Section 4.1.\footnote{If intra-national trade costs are modeled as an iceberg cost, denoted by $\widetilde{\tau} \ge 1$, the price index is modified as $P = [N (\widetilde{\tau} p)^{1-\sigma}]^{\frac{1}{1-\sigma}} = (\widetilde{\tau} a)^{\frac{1}{1-\mu}} N^{\frac{1}{(1-\sigma)(1-\mu)}}$ from Eqs. (3) and (7).
Since $\widetilde{\tau}$ and $a$ enter in a multiplicative form, both terms have the same effect on the $\Pi=0$-locus and the $\Delta r_m=0$-locus (see Eqs. (12) and (13)).
An increase in $\widetilde{\tau}$ as a result of a natural disaster shifts the $\Pi=0$-locus upward and the $N=N_0$-line rightward.
If the increase is so substantial that the $N=N_0$-line goes beyond the $N=\overline{N}$-line, the equilibrium switches from point $S_1$ to $S_2'$ (Proposition 2(i)).
A higher $\mu_m$ and a lower $\tau$ would move the $N=N_0$-line left (see Eq. (13)) and hence make the switch less likely to occur (Proposition 2(ii)).}
Our main results would maintain in this case.
That is, if the level of shock is substantial, the equilibrium switches from the one with both local and multinational firms to the one with only local firms (Proposition 2(i)).
Multinationals are more likely to stay in the disaster-hit host country if they rely more on local intermediates or they face lower trade costs (Proposition 2(ii)).

A more straightforward way to model natural disasters is direct shocks to multinationals, as documented in a number of disaster experiences in developing countries. 
Supposing that the fixed cost of capital for setting up a multinational increases from unity to $F_m>1$, other things being equal, the capital return generated by multinationals decreases and foreign capital no longer finds the host country profitable.
This results in a rightward shift of the $\Delta r_m=0$-locus.\footnote{
As a result of the disaster, the cost function of a typical multinational is modified as $C(q_m) = P^{\mu_m} (\tau p_u^*)^{1-\mu_m} \widetilde{a}_m q_m + r_m F_m$
and the capital return as $r_m = p_m q_m/(\sigma F_m)$.
The $\Delta r_m=0$-locus after shock is thus given by $N_0 = a^{\sigma-1} \left[ \tau^{\frac{1-\mu_m}{\mu_m}} (p_u^*)^{-1} \right]^{(\sigma-1)(1-\mu)} F_m^{\frac{1-\mu}{\mu_m}}$.
It can be seen that an increase in $F_m$ moves the $\Delta r_m=0$-locus right and a higher $\mu_m$ and a lower $\tau$ make the rightward shift smaller, leading to qualitatively the same results as Proposition 2.}
Applying an analogous reasoning, we can confirm qualitatively the same results as Proposition 2.

\section{Extensions}

\subsection{Multinationals with heterogeneous productivity}

In the basic model, multinationals are all homogeneous in productivity.
In reality, however, productivity differs among multinationals, which may lead to heterogeneous responses to disasters.
For example, \cite{Hayakawaetal2015} report that in the aftermath of the Thailand 2011 floods, changes in sourcing patterns of multinationals differ by their age and size, which are often associated with their productivity.
We will see that introducing heterogeneous productive multinationals into the basic model would give richer implications and more realistic equilibrium configurations.

Heterogeneity comes from a stochastic draw of unit input requirements, $\widetilde{a}_m = a_m (\sigma-1)/\sigma$, as in the literature on heterogeneous firm models (\citealp{Melitz2003}; \citealp{BaldwinOkubo2006}).
We assume that the cumulative density function of $a_m \in [\underline{a}_m, \overline{a}_m]$ takes a truncated Pareto distribution.
Namely, the probability of a multinational drawing productivity lower than $a_m$ is
\begin{align*}
&G(a_m) = \frac{ a_m^{\rho} - \underline{a}_m^{\rho} }{ \overline{a}_m^{\rho} - \underline{a}_m^{\rho} }
= \frac{ a_m^{\rho} - 1 }{  \overline{a}_m^{\rho} - 1 }, 
\end{align*}
where $\rho \ge 1$ is the shape parameter.
We set $\underline{a}_m$ to one without loss of generality.
In addition, it is reasonable to assume that high-productive MNEs (lower $a_m$) enjoy lower trade costs (lower $\tau$) per unit shipped than low-productive firms due to scale economies in transportation (\citealp{ForslidOkubo2015, ForslidOkubo2016}):\footnote{\cite{ForslidOkubo2016} document that the export-to-sales ratio of Japanese firms differs greatly among firms and is systematically higher for larger exporters.}
\begin{align*}
\tau(a_m) = a_m^{\gamma},
\end{align*}
where $\gamma>0$ governs the degree of scale economies in transportation.
This specification implies that most productive firms with $a_m=1$ have zero trade costs: $\tau(1)=1$.

The return to foreign capital depends on its productivity and thus its relocation incentive also differs by productivity.
Given the number of host domestic firms $N$, Fig. 5 shows a typical capital-return differential: $\Delta r_m(a_m) \equiv r_m(a_m) - r_f(a_m)$, which captures the incentive to locate in the host country.
The advantage of locating in the host country is a low cost of sourcing local intermediates, whereas its disadvantage is a high cost of sourcing foreign intermediates.
For multinationals with lower $a_m$, the advantage is more beneficial because of their larger amount of production, and the disadvantage is smaller thanks to scale economies of transportation.
Hence more productive multinationals have a greater $\Delta r_m(a_m)$ and thus a stronger incentive to locate in the host country, as shown in Fig. 5.

\

\begin{center}
\includegraphics[scale=1.0]{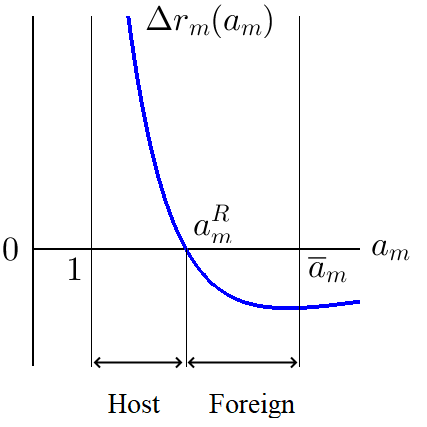} \\
Fig. 5. \ Location incentives of multinationals
\end{center}

\

As in the basic model, we consider a gradual relocation process.
Namely, each foreign capital, initially located in the foreign country, incurs migration costs $\chi$ to move to the host country.
The migration costs are assumed to be proportional to the flow of migrating capital, i.e., $\chi=\dot{N}_m$ (\citealp{BaldwinOkubo2006}).
Fig. 5 tells that the first capital ready to pay the relocation costs will be those that have the most to gain, i.e., the most efficient foreign capital.
In industry equilibrium where capital movement ceases, foreign capital with $a_m \le a_m^R$ enters in the host country, whereas foreign capital with $a_m > a_m^R$ remains in the foreign country, where $a_m^R$ is the cut-off productivity at which the two locations are indifferent.
The number of multinationals is thus given by
\begin{align*}
N_m = G(a_m^R)K_f = \frac{ (a_m^R)^{\rho} - 1  }{  \overline{a}_m^{\rho} - 1 } \cdot K_f,
\end{align*}
which increases $a_m^R$.

Taking the migration of foreign capital into account, the entry and exit of local suppliers gradually take place.
Fig. 6 illustrates industry equilibria in the $(N, N_m)$ plane, corresponding to Fig. 3.
Unlike the basic model, the $\Delta r_m(a_m^R)=0$-locus is no longer a vertical line.
Its upward slope comes from heterogeneous productivity: as the local supplying industry expands and the local-sourcing cost is lower, less productive foreign capital can enter.
As a result, a few but very productive multinationals remain in point $S_2$.
In Appendix 3, we can check that multiple equilibria arise if the fixed labor requirement for host domestic firms, $F$, takes an intermediate value, just like the basic model (Proposition 1(iii)).

\

\begin{center}
\includegraphics[scale=0.9]{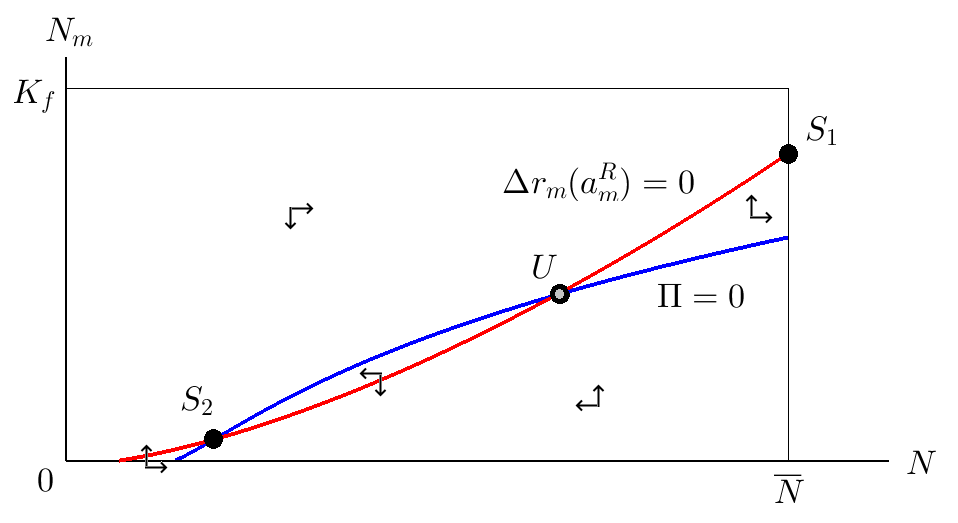} \\
Fig. 6. \ Industrial configurations under heterogeneous multinationals
\end{center}

\

Considering the effect of an increase in $F$ due to a natural disaster, we can see in Fig. 7 qualitatively the same changes and the same intuition behind them as in the basic model.
That is, the $\Pi=0$-locus moves up simply because lower profitability shrinks the area where host domestic firms can survive.
The total number of local firms that the host country can accommodate decreases so that the $N=\overline{N}$-line moves left.
As a result of these shifts, the new $\Pi'=0$-locus may no longer intersect at point $U$ with the $\Delta r_m(a_m^R)=0$-locus, in which case the equilibrium would switch from $S_1$ with many multinationals to $S_2'$ with few multinationals.
The unique but interesting difference from the basic model is clear from the new equilibrium $S_2'$: the most efficient multinationals are the ones that remain in the disaster-hit country.\footnote{This relocation pattern holds also when the disaster does not trigger the equilibrium switch. 
An increase in $F$ moves the $N=\overline{N}$-line left, thereby decreasing $N_m$ at equilibrium $S_1$ along the $\Delta r_m(a_m^R)=0$-locus.
A decrease in $N_m = G(a_m^R)K_f$ is equivalent with a decrease in $a_m^R$, implying that the least efficient multinationals are the first to leave the host country.}

\

\begin{center}
\includegraphics[scale=0.7]{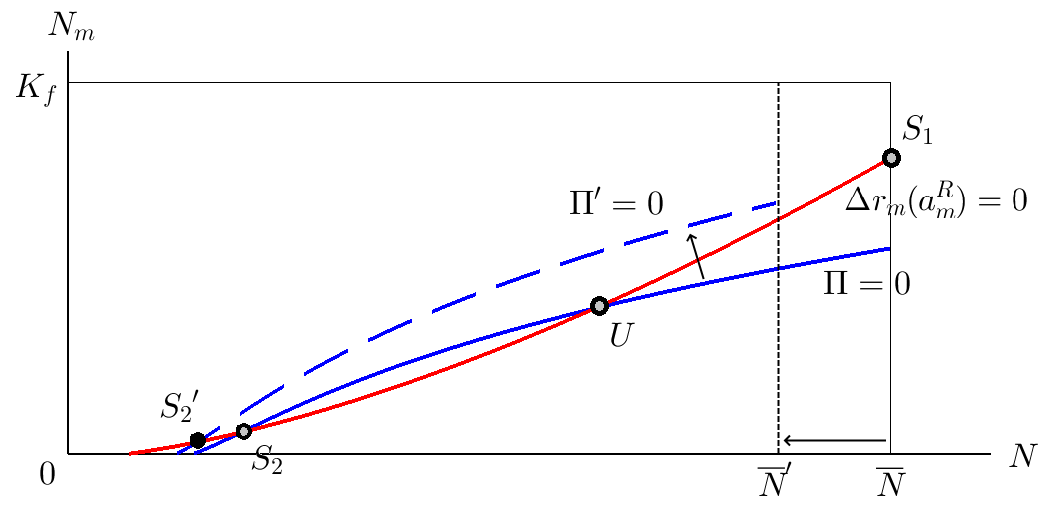} \\
Fig. 7. \ Impact of a natural disaster under heterogeneous multinationals
\end{center}

\

It can also be checked that if multinationals are more dependent on local suppliers (higher $\mu_m$), the disaster is less likely to trigger the equilibrium switch.\footnote{This result comes from the fact that a higher $\mu_m$ makes smaller the upward shift of the $\Pi=0$-locus.} 
By imposing assumptions similar to those in Propositions 1 and 2, we can formally prove the following proposition.
The proof and explicit expressions are given in Appendix 3.

\pagebreak

\noindent
{\bf Proposition 3 (Heterogeneous multinationals).} \ \ {\it Assume that multinationals are heterogeneous in a way that is specified in the text. 
Suppose that the fixed labor input, $F$, is intermediate so that multiple equilibria arise, i.e., $F \in (\widetilde{F}_a, \widetilde{F}_b]$, and that the host economy is initially at point $S_1$ with many multinationals.
Consider an increase in $F$ due to a natural disaster, then the following holds:
\begin{itemize}
\item[(i)] Multinationals with higher productivity are more likely to stay in the host country.
\item[(ii)] The natural disaster is less likely to trigger the equilibrium switch from $S_1$ to $S_2'$, where the host country loses most of multinationals, if they are more dependent on local intermediates (higher $\mu_m$).
\end{itemize}
}

\subsection{Reconstruction from disasters}

We have so far modeled a disaster as a permanent shock. 
Let us instead consider a temporal shock and uncover under what conditions foreign capital reenters the gradually recovering host country. 
The bottom line is that the host country re-attracts multinationals earlier, as (a)it recovers from the disaster more quickly, (b)they are more dependent on local intermediates or (c)they face lower trade costs.
These conditions have in common with those in Propositions 2(ii) and 3(ii).

As long as the myopic relocation decision is assumed as before, 
no foreign capital comes back to the host country even if it fully recovers from the disaster.
To allow for the possibility of reentering, we need to consider a forward-looking decision making of foreign capital: it chooses whether to become the multinational or the foreign domestic firm to maximize its lifetime return.\footnote{
We note that under the forward-looking behavior multiple equilibria disappear.
If the fixed labor input $F$ takes a value in between $F_a$ and $F_b$ and the situation is like Fig. 3, foreign capital chooses point $S_1$ at which it earns a higher flow return than at $S_2$, so that $S_1$ is the unique stable equilibrium.
Host domestic firms also prefer $S_1$ to $S_2$ because they make positive profits at $S_1$ but zero profits at $S_2$.
Point $S_1$ is better than $S_2$ in the Pareto sense.}

Consider the situation where the host country is hit by a natural disaster at time $s=0$ and all multinationals leave there, corresponding to point $S_2'$ in Fig. 4(c).
Suppose then that the increased fixed-labor input due to the disaster gradually gets back to the pre-shock level after some time $T$.
Specifically, the fixed labor input at time $s$, denoted by $F(s)$, is given by
\begin{align*}
F(s) =\begin{cases}
Fe^{\delta(T-s)} &\text{for} \ \ s \in [0, T) \\
F &\text{for} \ \ s \in [T, \infty)
\end{cases},
\end{align*}
where $\delta$ is the recovery rate and we assume $F(0) = Fe^{\delta T} > F_b \equiv L/N_0[\sigma(1-\mu) +\mu]$ and $F \le F_b$.
Letting $t \ (\le T)$ be the time at which foreign capital moves from the foreign to the host country, the lifetime capital return, $v(t)$, is given by
\begin{align*}
v(t) &\equiv \underbrace{\int_0^t e^{-\theta s} r_f ds}_{\text{Locate in foreign}} + \underbrace{\int_t^{T} e^{-\theta s} r_m'(s) ds
+\int_T^{\infty} e^{-\theta s} r_m ds}_{\text{Locate in host}},
\end{align*} 
where $\theta>0$ is the discount rate.
For $s \in [0, t)$, foreign capital remains in the foreign country and makes the flow return $r_f$.
It locates in the host country from time $t$ onward.
The flow return during recovery period $s \in [t, T)$ is $r_m'(s)$ and is fully recovered to the pre-shock level $r_m (\ge r_m'(s))$ for $s \in [T, \infty)$.
The optimal timing of relocation can be derived from maximizing $v(t)$ with respect to $t$.

One naturally expects that if reconstruction takes very long time (large $T$), foreign capital with positive discount rate $\theta$ never relocates to the host country, i.e., $t=\infty$.
Furthermore, we see from Section 4.1 that multinationals make greater profits as the cost share of local intermediate goods in multinational production, $\mu_m$, is higher or trade costs, $\tau$, are lower.
This should imply that a higher $\mu_m$ or a lower $\tau$ would make multinationals' reentering earlier (lower $t$).
Assuming (a) $\alpha>1-(1-\mu)/\mu_m$ and (b) $K_f > \Theta \overline{N}^{\frac{1-\mu-\mu_m}{1-\mu}} (1-\alpha)$ as in Propositions 1 and 2, we can establish these arguments in Proposition 4 and formally prove them in Appendix 4.

\

\noindent
{\bf Proposition 4 (Recovery from disaster).} \ \ {\it Consider the situation where a natural disaster hits the host country and there are no multinationals at time $s=0$.
The fixed labor input at time $s=0$ is given by $F(0) = Fe^T$ and gradually recovers to the pre-shock level $F$, according to $F(s)=Fe^{\delta (T-s)}$, where it takes time $T$ to fully recover and we assume $F \le F_b < F(0)=Fe^T$.
On the optimal timing $s=t$ of multinational reentering, the following holds:
\begin{itemize}
\item[(i)] Multinationals never reenter the host country, i.e., $t=\infty$, if the recovery time from the disaster is too long such that $T >> \overline{T}$, where $\overline{T}$ is a bundle of parameters distinct from $T$.
\item[(ii)] Assuming a range of parameters that ensure interior solutions of optimal timing $t \in (0, \infty)$, multinationals reenter the host country more quickly (smaller $t$), as the recovery time is shorter (smaller $T$).
Assume, in addition, that trade costs are positive ($\tau > 1$) and the recovery time $T$ is greater than but close to a bundle of parameters $\widehat{T}$ distinct from $T$.
Then, the optimal reentering timing $t$ decreases with the cost share of local intermediate goods in multinational production ($\mu_m$) and increases with trade costs ($\tau$).
\end{itemize}
}

\

\subsection{Other extensions}

\subsubsection{Host country's market}

In the basic model, we assume that multinationals do not serve consumers in the host country.
We can extend it in a way such that multinationals can provide final goods with both the host and the foreign markets, while foreign domestic firms still provide with the foreign market only.
In the extended model, foreign capital is more likely to enter the host country rather than to remain in the foreign country.
In fact, we can confirm that in any equilibria there is a positive number of multinationals.
See Appendix 5 for details.

\subsubsection{Endogenous sourcing patterns}

In general, natural disasters do not only affect the location choice of multinationals but also their linkages with local suppliers.
To describe the endogenous sourcing patterns, we allow foreign capital to choose either of the two types of multinationals: one type, called the $H$-multinational, with a high cost share of local intermediate goods $\mu_m^H \in (0, 1)$ and the other type, called the $L$-multinational, with a low cost share $\mu_m^L (<\mu_m^H)$. 
Because building a close relationship with local suppliers needs huge investment, it is reasonable to assume that the $H$-multinational incurs a greater fixed capital input than the $L$-multinational.
Consider then a natural disaster hitting the host country where the $H$-multinationals operate, raising the fixed labor input for local suppliers $F$.
If the magnitude of the shock is moderate, multinationals continue to stay in the host country, but they all switch to the $L$-multinational and depend less on local intermediates.
The damaged local industry decreases the profitability of local sourcing and is thus unable to support the $H$-multinationals, which need higher fixed setup costs.
See Appendix 6 for details.

\subsubsection{Disaster risk and the timing of leaving}

Although the basic model assumes no uncertainty and myopic location decisions, empirical findings suggest that investor perceptions of risk of future disasters may also shape investment decisions in the long run (\citealp{TonerFriedt2020}).
The disaster risk can be incorporated into the model in a way such that forward-looking multinationals choose the timing of leaving the host country while expecting future disasters.
We can show that even when the damage of the (initial) shock is small, uncertainty itself may lead multinationals to move out of the host country even before the actual shock hits.
The duration of stay becomes shorter as they perceive a higher frequency of disaster occurrence.
See Appendix 7 for details.

\

\section{Conclusion}

This paper has developed a theoretical framework to address the resilience of multinationals against a severe shock such as natural disasters.
Our focus is on two notable aspects of multinationals, footloose-ness and input-output linkages with local suppliers.
These aspects give rise to multiple equilibria, one in which multinationals help local industry develop and the other in which they never enter.
When a natural disaster seriously damages local suppliers and thus raises the prices of local intermediates, the equilibrium switch occurs: multinationals leave the host country and shall never return.
The extreme case of the complete exit of multinationals can be avoided when allowing for heterogeneous productivity and the host country's market.
The discontinuous change in equilibrium due to severe disasters may help explain mixed evidences on the disaster impact on FDI mentioned in Introduction.
When using disaster variables pooling all events with different intensities of damage, one may fail to find statistically significant coefficients on them.
Mixed results may also come from sample countries with different development levels.
In addition, since the disaster preparedness of a country typically depends on its economic development level, damages are more likely to be severe in developing countries than developed ones.

Using the framework, we have identified under what conditions multinationals are more likely to stay in the disaster-hit host country.
The key parameters are the share of local intermediate goods in multinational production and trade costs of foreign intermediate goods.
In particular, as multinationals rely more on local suppliers and make greater profits through low sourcing costs, a decline in the local supplying industry due to natural disasters is less likely to affect their relocation decision.
This insight carries over to the case where multinationals are heterogeneous and to the analysis of the timing of disaster reconstruction.

We believe that our model yields rich analytical outcomes, yet remains sufficiently simple to produce new insights into the nexus between natural disasters and multinationals.
The analysis can be enriched in many ways.
One way is to explicitly introduce local upstream and downstream firms and allow them to benefit from technology spillovers from multinationals.
Using the extended model, one can distinguish between inter-industry (i.e., horizontal) and intra-industry (i.e., vertical) spillovers and examine their interactions with natural disasters.
The degree of spillovers may decrease when greater disaster risk discourages MNEs' commitment to local procurement. 
We leave this and other possible extensions to future research.

\ 

\ 

%\section*{Acknowledgements}
 
\begin{spacing}{1}

\section*{Appendices}

\section*{Theory Appendix}

\subsection*{Appendix 1. \ \ Derivations}

We here provide detailed derivations of cost function $C(q)$, total demand for a differentiated product $q$, and free entry conditions.

\subsubsection*{A1-1. \ \ Cost function}

The cost-minimization problem for a typical domestic firm producing variety $\omega$ is 
\begin{align*}
&\min_{ \{ q_u(\omega')\}, l } \int p(\omega') q_u(\omega')d\omega' + wl + w F, \\
\
&\text{s.t.} \ \ \widetilde{a} q(\omega) = z \left[  \left( \int q_u(\omega')^{\frac{\sigma -1}{\sigma }}d\omega'\right)^{\frac{\sigma}{\sigma-1}} \right]^{\mu} l^{1-\mu},
\end{align*}
where $q_u(\omega')$ is intermediate demand for variety $\omega'$.
The symmetry of firms implies $q(\omega) = q$.
The problem can be solved in two steps.
First, we consider the cost-minimization problem for the differentiated inputs:
\begin{align*}
&\min_{ \{ q_u(\omega')\} } \int p(\omega') q_u(\omega')d\omega', \\
\
&\text{s.t.} \ \ Q_u = \left( \int q_u(\omega')^{\frac{\sigma -1}{\sigma }}d\omega' \right)^{\frac{\sigma}{\sigma-1}}.
\end{align*}
The FOCs yield $q_u(\omega') = [p(\omega')/P]^{-\sigma} Q_u$.
Using this, we reformulate the original problem as
\begin{align*}
&\min_{ Q_u, l } P Q_u + w l + w F, \\
\
&\text{s.t.} \ \ \widetilde{a} q = z Q_u^{\mu} l^{1-\mu},
\end{align*}
The FOCs for the above minimization problem give demand functions for the composite input and labor:
\begin{align*}
&Q_u = \mu z \mu^{-\mu} (1-\mu)^{-(1-\mu)} P^{\mu-1} w^{1-\mu} \widetilde{a} q, \\
\
&l =  (1-\mu) z \mu^{-\mu} (1-\mu)^{-(1-\mu)} P^{\mu} w^{-\mu} \widetilde{a} q.
\end{align*}
Substituting these results back into the objective function and choosing $z$ as $z = \mu^{\mu} (1-\mu)^{1-\mu}$, we obtain the cost function for the host domestic firm in the main text.
The cost function for the multinational is obtained in a similar way.

%{\it Total demand.} \ \ 
\subsubsection*{A1-2. \ \ Total demand}

We apply Shephard's lemma to the cost function given in Eq. (5) to obtain the intermediate-good demand by host domestic firms for variety $\omega$:
\begin{align*}
\frac{\partial C(q)}{\partial p(\omega')} &= \frac{\partial P^{\mu}}{\partial p(\omega')} \cdot w^{1-\mu} \widetilde{a} q \\
\
&= \frac{\partial }{\partial p(\omega')} \left( \int p(\omega')^{1-\sigma} d\omega \right)^{ \frac{\mu}{1-\sigma} } \cdot \widetilde{a} q \\
\
&= \frac{\mu}{1-\sigma} (1-\sigma) p(\omega')^{-\sigma} P^{\mu -1} \widetilde{a} q \\
\
&= \mu p(\omega')^{-\sigma} P^{\sigma +\mu -1} \widetilde{a} q,
\end{align*}
noting that $w=1$.
As all host domestic firms are symmetric, their input demand for the variety becomes $N \mu p^{-\sigma} P^{\sigma +\mu -1} \widetilde{a} q$.
Similarly, we derive the intermediate-good demand by all multinationals as $N_m \mu_m p_m^{-\sigma} P^{\sigma +\mu_m -1} (\tau p_u^*)^{1-\mu_m} \widetilde{a}_m q_m$.
The total demand for the variety is the sum of these intermediate-good demand and final-good demand (Eq.(2)):
\begin{align*}
q &= \left( \frac{p}{P} \right)^{-\sigma} \frac{E}{P} + N \mu p^{-\sigma} P^{\sigma +\mu -1} w^{1-\mu}  \widetilde{a} q + N_m \mu_m p^{-\sigma} P^{\sigma +\mu_m -1} (\tau p_u^*)^{1-\mu_m} \widetilde{a}_m q_m \\
\
&= p^{-\sigma} \left[ P^{\sigma-1}E  + N \mu P^{\sigma +\mu -1} \widetilde{a} q + N_m \mu_m P^{\sigma +\mu_m -1} (\tau p_u^*)^{1-\mu_m} \widetilde{a}_m q_m \right]. \tag{11}
\end{align*}
which is given by Eq. (11) in the main text.

\subsubsection*{A1-3. \ \ Free entry conditions}

The following expressions are useful for later reference:
\begin{align*}
&P = a^{\frac{1}{1-\mu}} N^{\frac{1}{(1-\sigma)(1-\mu)}}, \tag{A1} \\
\
&p^{1-\sigma} = \left( aP^{\mu} \right)^{1-\sigma} =a^{\frac{1-\sigma}{1-\mu}} N^{\frac{1}{1-\mu}}, \tag{A2} \\
\
&p_m^{1-\sigma} = \left[ a_m P^{\mu_m} (\tau p_u^*)^{1-\mu_m} \right]^{1-\sigma} 
= \left[ a_m a^{\frac{\mu_m}{1-\mu}}  (\tau p_u^*)^{1-\mu_m} \right]^{1-\sigma} N^{\frac{\mu_m}{1-\mu}}. \tag{A3}
\end{align*}

Free entry and exit imply that no host domestic firms enter if their excess profits are negative, $\Pi = pq-C(q)<0$, while if $\Pi \ge 0$ there are positive entries.
A typical domestic firm breaks even, if 
\begin{align*}
0 = \Pi &= pq - C(q) \\
\
&= pq - P^{\mu} \widetilde{a} q - F \\
\
&= pq - \frac{\sigma-1}{\sigma} \frac{\sigma P^{\mu} \widetilde{a} }{\sigma-1} q - F \\
\
&= pq - \frac{\sigma-1}{\sigma} pq - F \\
\
&= pq/\sigma -F,
\end{align*}
where we used Eq. (6) from the third to the fourth line.
The break-even level of sales of host domestic firms are thus $pq = \sigma F$.
They make positive excess profits if the differentiated sector uses up local labor and thus no further entry into the sector is possible.

We multiply both sides of Eq. (11) by $p$ to get
\begin{align*}
pq &= \left( \frac{p}{P} \right)^{1-\sigma} E + N \mu \left( \frac{p}{P} \right)^{1-\sigma} P^{\mu} \widetilde{a} q + N_m \mu_m  \left( \frac{p}{P} \right)^{1-\sigma} P^{\mu_m} (\tau p_u^*)^{1-\mu_m} \widetilde{a}_m q_m \\
\
&= \frac{E}{N} + \mu \frac{\sigma-1}{\sigma} \frac{\sigma \widetilde{a} P^{\mu} }{\sigma-1} q
+ \mu_m \frac{N_m}{N} \frac{\sigma-1}{\sigma} \frac{\sigma \widetilde{a}_m P^{\mu_m} (\tau p_u^*)^{1-\mu_m}}{\sigma-1} q_m \\
\
&= \frac{E}{N} + \frac{\sigma-1}{\sigma} \mu pq + \frac{\sigma-1}{\sigma} \mu_m \frac{N_m}{N} p_m q_m,
\end{align*}
where we used Eq. (3) from the first to the second line and Eqs. (6) and (9) from the second to the last line.
Solving this equation for $pq$ gives
\begin{align*}
pq &= \frac{\sigma}{\sigma - \mu(\sigma-1)} \left[ \frac{E}{N} + \frac{\mu_m (\sigma-1)}{\sigma} \frac{N_m}{N} p_m q_m \right] \\
\
&= \frac{\sigma}{\sigma(1-\mu) + \mu} \left[ \frac{E}{N} + \frac{\mu_m (\sigma-1)}{\sigma} \frac{N_m}{N} p_m^{1-\sigma} D^* \right].
\end{align*}
When there are no excess profits for host domestic firms, the aggregate expenditure on differentiated goods is a $\alpha$ share of total labor income: $E = \alpha w L = \alpha L$.
Substituting the above expression, Eq. (A3) and $E=\alpha L$ into the break-even level of sales yields
\begin{align*}
&pq = \sigma F, \\
\
&\to \frac{\sigma}{\sigma(1-\mu) + \mu} \left[ \frac{\alpha L}{N} + \frac{\mu_m (\sigma-1)}{\sigma} \frac{N_m}{N}  N^{\frac{\mu_m}{1-\mu}} \left\{ a_m a^{\frac{\mu_m}{1-\mu}}  (\tau p_u^*)^{1-\mu_m} \right\}^{1-\sigma} D^*  \right] = \sigma F, \\
\
&\to \alpha L + \frac{\mu_m D^* (\sigma-1)}{\sigma} \left[ a_m a^{\frac{\mu_m}{1-\mu}}  (\tau p_u^*)^{1-\mu_m} \right]^{1-\sigma} N^{\frac{\mu_m}{1-\mu}} N_m = N F [\sigma(1-\mu) + \mu], \\
\
&\to N_m = \Theta N^{-\frac{\mu_m}{1-\mu}}  \left( N  -\alpha \overline{N} \right), \tag{12} \\
\
&\text{where} \ \ \Theta \equiv \frac{ \sigma F  [\sigma(1-\mu)+\mu] \left[ a_m a^{\frac{\mu_m}{1-\mu}}  (\tau p_u^*)^{1-\mu_m} \right]^{\sigma-1} }{\mu_m D^*(\sigma-1) }, \ \ \ \
\overline{N} \equiv \frac{L}{F[\sigma(1-\mu)+\mu]},
\end{align*}
which is given by Eq. (12) in the main text.

Free entry and exit of multinationals and foreign domestic firms drive their excess profits to zero, which determines their rental rate of capital:
\begin{align*}
&\Pi_m = p_m q_m - C_m(q_m) = p_m q_m - P^{\mu_m} (\tau p_u^*)^{1-\mu_m} \widetilde{a}_m q_m - r_m = 0, \\
\
&\to r_m = p_m q_m/\sigma = p_m^{1-\sigma} D^*/\sigma, \\
\
&\Pi_f = p_f q_f - C_f(q_f) = p_f q_f -  p_u^* \widetilde{a}_m q_f- r_f  = 0, \\
\
&\to r_f = p_f q_f/\sigma = p_f^{1-\sigma} D^*/\sigma.
\end{align*}
Foreign capital is indifferent between becoming a multinational and a foreign domestic firm if the return differential is zero:
\begin{align*}
\Delta r_m \equiv r_m -r_f &= D^* (p_m^{1-\sigma}  - p_f^{1-\sigma} )/\sigma \\
\
&= D^* \left[ \left( a_m P^{\mu_m} (\tau p_u^*)^{1- \mu_m} \right)^{1-\sigma} - (a_m p_u^*)^{1-\sigma} \right]/\sigma \\
\
&= D^*  \left[  \left\{ a_m a^{\frac{\mu_m}{1-\mu}}  (\tau p_u^*)^{1-\mu_m} \right\}^{1-\sigma} N^{\frac{\mu_m}{1-\mu}}-  (a_m p_u^*)^{1-\sigma}  \right]/\sigma \\
\
&= D^* (a_m p_u^*)^{1- \sigma} \left[  \left\{ a^{\frac{\mu_m}{1-\mu}} \tau^{1-\mu_m} (p_u^*)^{-\mu_m} \right\}^{1-\sigma} N^{\frac{\mu_m}{1-\mu}} -  1  \right]/\sigma = 0. \tag{A4}
\end{align*}
Solving this for $N$ gives
\begin{align*}
N =  a^{\sigma-1} \left[ \tau^{\frac{1-\mu_m}{\mu_m}} (p_u^*)^{-1} \right]^{(\sigma-1)(1-\mu)}  \equiv N_0, \tag{13}
\end{align*} 
which is given by Eq. (13) in the main text.

\subsubsection*{A1-4. \ \ Upper bound of the number of host domestic firms}

Aggregate labor demand in the differentiated sector of the host country is the product of the labor demand by individual host domestic firms and their total number, $N$.
Applying Shephard's lemma to Eq. (5), we obtain
\begin{align*}
N \frac{\partial C(q)}{\partial w} &= N \frac{\partial ( P^{\mu} w^{1-\mu} \widetilde{a}q +wF)}{\partial w} \\
\
&= N \left[ (1-\mu) P^{\mu} w^{-\mu} \widetilde{a}q + F \right] \\
\
&= N \left[ (1-\mu) P^{\mu} \widetilde{a}q + F \right].
\end{align*}
We evaluate this at the break-even level of sales, $pq=\sigma F$, to get
\begin{align*}
N \left[ (1-\mu) P^{\mu} \widetilde{a}q + F \right] &=  N \left[ (1-\mu)(\sigma-1)pq/\sigma + F \right] \\
\
&= N \left[ (1-\mu)(\sigma-1)\sigma F/\sigma + F \right] \\
\
&= N F [\sigma(1-\mu) + \mu].
\end{align*}
The labor demand must be smaller than the total workforce in the host country, $L$:
\begin{align*}
&N F [\sigma(1-\mu) + \mu] \le L, \\
\
&\to N \le \frac{L}{F[\sigma(1-\mu) + \mu]} \equiv \overline{N},
\end{align*}
which determines the upper bound of $N$.

\

\subsection*{Appendix 2. \ \ Impact of natural disaster}

\subsection*{A2-1. \ \ Proof of Proposition 2}

Since Proposition 2(i) is established in the main text, here we prove Proposition 2(ii): $\partial (\Delta F_{min})/\partial \mu_m \ge 0$; $\partial (\Delta F_{min})/\partial \tau \le 0$.
Differentiating $\Delta F_{min}$ defined in Eq. (14) with respect to $\mu_m$ yields
\begin{align*}
\frac{\partial (\Delta F_{min})}{\partial \mu_m} &= \frac{\partial }{\partial \mu_m} \left[ \frac{L}{N_0 \{ \sigma(1-\mu) + \mu \}} - F \right] \\
\
&= -\frac{L}{N_0^2 [\sigma(1-\mu) + \mu]} \cdot \frac{\partial N_0}{\partial \mu_m} \\
\
&= -\frac{L}{N_0^2 [\sigma(1-\mu) + \mu]} \cdot \frac{\partial}{\partial \mu_m} \left[ a^{\sigma-1} \left\{  \tau^{\frac{1-\mu_m}{\mu_m}} (p_u^*)^{-1} \right\}^{(\sigma-1)(1-\mu)} \right] \\
\
&= -\frac{L}{N_0^2 [\sigma(1-\mu) + \mu]} \cdot N_0 \left[ -\frac{(\sigma-1)(1-\mu)}{\mu_m^2} \ln \tau \right] \\
\
&= \frac{L (\sigma-1)(1-\mu) \ln \tau }{\mu_m^2 \tau N_0 [\sigma(1-\mu) + \mu]} \ge 0,
\end{align*}
where equality holds at $\tau=1$.

Similarly, differentiating $\Delta F_{min}$ with respect to $\tau$ gives
\begin{align*}
\frac{\partial (\Delta F_{min})}{\partial \tau} &= -\frac{L}{N_0^2 [\sigma(1-\mu) + \mu]} \cdot \frac{\partial N_0}{\partial \tau} \\
\
&= -\frac{L}{N_0^2 [\sigma(1-\mu) + \mu]} \cdot \frac{N_0}{\tau} \cdot \frac{(\sigma-1)(1-\mu)(1-\mu_m)}{\mu_m} \\
\
&= -\frac{L (\sigma-1)(1-\mu)(1-\mu_m) \ln \tau }{\mu_m N_0 [\sigma(1-\mu) + \mu]} \le 0,
\end{align*}
where equality holds at $\tau=1$.
These establish Proposition 2(ii).

\

\subsection*{A2-2. \ \ Other issues}

We derive the condition for $\partial (\Delta F_{min})/\partial \mu > 0$ if $F \in (F_a, F_b)$:
\begin{align*}
\frac{\partial (\Delta F_{min})}{\partial \mu} &= \frac{\partial }{\partial \mu} \left[ \frac{L}{N_0 \{ \sigma(1-\mu) + \mu \}} - F \right] \\
\
&=  \frac{L(\sigma-1)}{N_0[\sigma(1-\mu) + \mu]^2} - \frac{L}{N_0^2 [\sigma(1-\mu) + \mu]} \frac{\partial N_0}{\partial \mu} \\
\
&= \frac{L(\sigma-1)}{N_0[\sigma(1-\mu) + \mu]} \left[ \frac{1}{\sigma(1-\mu) + \mu} + \ln \left\{ \tau^{\frac{1-\mu_m}{\mu_m}} (p_u^*)^{-1} \right\} \right] \\
\
&= \frac{F \overline{N} (\sigma-1)}{N_0} \left[ \frac{1}{\sigma(1-\mu) + \mu} + \ln \left\{ \tau^{\frac{1-\mu_m}{\mu_m}} (p_u^*)^{-1} \right\} \right] \\
\
&> F(\sigma-1) \left[ \frac{1}{\sigma(1-\mu) + \mu} + \ln \left\{ \tau^{\frac{1-\mu_m}{\mu_m}} (p_u^*)^{-1} \right\} \right] \equiv  RHS(\mu).
\end{align*}
where from the second-to-last to the last line we used the fact that $F < F_b$, or equivalently $N_0 < \overline{N}$.
A sufficient condition for $\partial (\Delta F_{min})/\partial \mu > 0$ is $RHS(\mu) \ge 0$ for $\mu \in (0, 1)$.
While noting that $RHS'(\mu)>0$ holds, we use the expression of $N_0$ given in Eq. (13) to rewrite $RHS(\mu)$ as
\begin{align*}
RHS(\mu) &=  F(\sigma-1) \left[ \frac{1}{\sigma(1-\mu) + \mu} + \frac{ \ln \left( a^{1-\sigma} N_0 \right) }{(\sigma-1)(1-\mu)} \right] \\
\
&> RHS(0) = F(\sigma-1) \left[ \frac{1}{\sigma} + \frac{\ln \left( a^{1-\sigma} N_0 \right) }{\sigma-1} \right] \\
\
&\ge F(\sigma-1) \left[ \frac{1}{\sigma} + \frac{\ln \left( a^{1-\sigma} \alpha \overline{N} \right) }{\sigma-1} \right].
\end{align*}
The expression in the last line is non-negative if
\begin{align*}
&\frac{1}{\sigma} + \frac{\ln \left( a^{1-\sigma} \alpha \overline{N} \right) }{\sigma-1} \ge 0, \\
\
&\to a \le e^{\frac{1}{\sigma}} (\alpha \overline{N})^{\frac{1}{\sigma-1}},
\end{align*}
which is a sufficient condition for $\partial (\Delta F_{min})/\partial \mu > 0$.

Next we show $P \le \tau p_u^*$ if there exists $N_0 \ge 0$ that satisfies Eq. (13) (or equivalently Eq. (A4)).
Let us first look at Eq. (A4):
\begin{align*}
&\Delta r_m = D^* (a_m p_u^*)^{1- \sigma} [ g(N; \mu_m) -  1 ]/\sigma = 0, \tag{A4} \\
\
&\text{where} \ \ g(N; \mu_m) \equiv \left[ a^{\frac{\mu_m}{1-\mu}} \tau^{1-\mu_m} (p_u^*)^{-\mu_m} \right]^{1-\sigma} N^{\frac{\mu_m}{1-\mu}} \ge 0.
\end{align*}
We focus on a range of parameters in which we can always find $N$ that satisfies $g(N; \mu_m)=1$ for any $\mu_m \in (0, 1)$.
Because $g(N; \mu_m=0) = \tau^{1-\sigma} \le 1$, it must be that $\partial g(N; \mu_m)/\partial \mu_m \ge 0$ at $\mu_m=0$; otherwise there would be no $N$ that satisfies $g(N; \mu_m)=1$
This condition results in $\tau p_u^*/P \ge 1$:
\begin{align*}
&\frac{\partial g(N; \mu_m)}{\partial \mu_m}\bigg|_{\mu_m=0} = g(N; 0) \cdot \ln \left[ \tau p_u^* \left( a N^{\frac{1}{1-\sigma}} \right)^{-\frac{1}{1-\mu}} \right]^{\sigma-1} \ge 0, \\
\
&\to \ln \left[ \tau p_u^* \left( a N^{\frac{1}{1-\sigma}} \right)^{-\frac{1}{1-\mu}} \right]^{\sigma-1} \ge 0, \\
\
&\to \tau p_u^* \left( a N^{\frac{1}{1-\sigma}} \right)^{-\frac{1}{1-\mu}} = \tau p_u^*/P \ge 1,
\end{align*}
noting that the derivative of $f(x) = a^{A(x)} b^{B(x)}$ is given by $f'(x) = f(x) \ln \left[ a^{A'(x)} b^{B'(x)} \right]$.

\

\subsection*{Appendix 3. \ \ Multinationals with heterogeneous productivity}

\subsubsection*{A3-1. \ \ Location condition of multinationals}

Using the results in Appendix A1-3, we can write the return differential of foreign capital as
\begin{align*}
\Delta r_m(a_m) &= r_m(a_m) - r_f(a_m) \\
\
&= p_m(a_m)q_m(a_m)/\sigma - p_f(a_m)q_f(a_m)/\sigma \\
\
&= D^* \left[ p_m(a_m)^{1-\sigma} - p_f(a_f)^{1-\sigma} \right]/\sigma \\
\
&= D^* (a_m p_u^*)^{1- \sigma} \left[  \left\{ a^{\frac{\mu_m}{1-\mu}} (a_m^{\gamma})^{1-\mu_m} (p_u^*)^{-\mu_m} \right\}^{1-\sigma} N^{\frac{\mu_m}{1-\mu}} -  1  \right]/\sigma,
\end{align*}
noting that $\tau(a_m) = a_m^{\gamma}$.
The cut-off productivity, $a_m^R$, is given by the solution of $\Delta r_m(a_m^R)=0$.
Noting that the big square bracket term in the above expression decreases with $a_m$ because $\gamma(1-\mu_m)(1-\sigma)<0$,
the return differential is positive (or negative) if $a_m < a_m^R$ (or $a_m > a_m^R$).

We can explicitly solve for the cut-off productivity:
\begin{align*}
&\Delta r_m(a_m^R) = 0, \\
\
&\to (a_m^R)^{\gamma(1-\mu_m)(1-\sigma)} 
\left\{ a^{\frac{\mu_m}{1-\mu}} (p_u^*)^{-\mu_m} \right\}^{1-\sigma} N^{\frac{\mu_m}{1-\mu}} -1 > 0, \\
\
&\to (a_m^R)^{\gamma(1-\mu_m)(1-\sigma)} = \left\{ a^{\frac{\mu_m}{1-\mu}}  (p_u^*)^{-\mu_m} \right\}^{-(1-\sigma)} N^{-\frac{\mu_m}{1-\mu}}, \\
\
&\to a_m^R = \left\{ a^{\frac{1}{1-\mu}}
 (p_u^*)^{-1} \right\}^{ - \frac{\mu_m}{\gamma(1-\mu_m)} } N^{ \frac{\mu_m}{\gamma(\sigma-1)(1-\mu)(1-\mu_m)} }.
\end{align*}

The number of multinationals is expressed as a function of the cut-off productivity: 
\begin{align*}
&N_m = \Pr(\widehat{a}_m \le a_m^R) = G(a_m^R)K_f
= \frac{K_f(a_m^{\rho} -1)}{\overline{a}_m^{\rho} -1}, \\
\
&\to a_m^R = [N_m ( \overline{a}_m^{\rho} -1 )/K_f  + 1]^{1/\rho}. \tag{A5}
\end{align*}
Substituting this into the explicit solution of $a_m^R$ gives
\begin{align*}
&[N_m ( \overline{a}_m^{\rho} -1 )/K_f  + 1]^{1/\rho}
= \left\{ a^{\frac{1}{1-\mu}} (p_u^*)^{-1} \right\}^{ - \frac{\mu_m}{\gamma(1-\mu_m)} } N^{ \frac{\rho \mu_m}{\gamma(\sigma-1)(1-\mu)(1-\mu_m)} }, \\
\
&\to N_m ( \overline{a}_m^{\rho} -1 )/K_f  + 1 
= \left\{ a^{\frac{1}{1-\mu}} (p_u^*)^{-1} \right\}^{ - \frac{\rho \mu_m}{\gamma(1-\mu_m)} } N^{ \frac{\rho \mu_m}{\gamma(\sigma-1)(1-\mu)(1-\mu_m)} }, \\
\
&\to N_m = \frac{K_f}{ \overline{a}_m^{\rho} -1 }
\left[ \left\{ a^{\frac{1}{1-\mu}} (p_u^*)^{-1} \right\}^{ - \frac{\rho \mu_m}{\gamma(1-\mu_m)} } N^{ \frac{\rho \mu_m}{\gamma(\sigma-1)(1-\mu)(1-\mu_m)} } - 1 \right], \tag{A6}
\end{align*}
which increases with $N$ because $\rho \mu_m/[\gamma(\sigma-1)(1-\mu)(1-\mu_m)]>0$.
Fig. A1(b) draws the $\Delta r_m(a_m^R)=0$-locus defined in Eq. (A6), where $\widetilde{N}_0$, at which $N_m=0$ holds, is given by
\begin{align*}
&N_m = 0, \\
\
&\to N = a^{\sigma-1} \left[(p_u^*)^{-1} \right]^{(\sigma-1)(1-\mu)} \equiv \widetilde{N}_0,
\end{align*}
which is assumed to be greater than one: $\widetilde{N}_0 > 1$. 
The arrows in Fig. A1(b) indicate the direction of motion of foreign capital.
As in the basic model, the area where foreign capital moves to the host country expands as $N$ increases.
Unlike the basic model, however, the locus is not a vertical line but a upward-sloping curve because the relocation incentive differs in productivity.
High-productive foreign capital is ready to move to the host country with a small number of suppliers, whereas a sufficient number of local suppliers is necessary for low-productive one to move.

\subsubsection*{A3-2. \ \ Zero-profit conditions of host domestic firms}

The goods market clearing condition requires that the total sales must be equal to the total purchase by consumers, host domestic firms and multinationals:
\begin{align*}
pq &= \left( \frac{p}{P} \right)^{1-\sigma} E + N \mu \left( \frac{p}{P} \right)^{1-\sigma} P^{\mu} \widetilde{a} q + K_f \mu_m \left( \frac{p}{P} \right)^{1-\sigma}  \int_1^{a_m^R} P^{\mu_m} [\tau(a_m) p_u^*]^{1-\mu_m} \widetilde{a}_m q_m(a_m) dG(a_m) \\
\
&= \frac{E}{N} + \mu \frac{\sigma-1}{\sigma} \frac{\sigma \widetilde{a} P^{\mu} }{\sigma-1} q
+ \mu_m \frac{K_f}{N} \frac{\sigma-1}{\sigma} \int_1^{a_m^R} \frac{\sigma \widetilde{a}_m P^{\mu_m} [\tau(a_m) p_u^*]^{1-\mu_m}}{\sigma-1} q_m(a_m) dG(a_m) \\
\
&= \frac{E}{N} + \frac{\sigma-1}{\sigma} \mu pq + \frac{\sigma-1}{\sigma} \mu_m \frac{K_f}{N} \int_1^{a_m^R} p_m(a_m) q_m(a_m)dG(a_m),
\end{align*}
noting that multinationals with $a_m \in [1, a_m^R]$ are in the host country.
The integral part in the right-hand side is 
\begin{align*}
\int_1^{a_m^R} p_m(a_m) q_m(a_m)dG(a_m) &= \int_1^{a_m^R} p_m(a_m)^{1-\sigma} D^* \cdot \frac{\rho a_m^{\rho-1}}{ \overline{a}_m^{\rho} -1 } da_m \\
\
&= \frac{\rho D^*}{ \overline{a}_m^{\rho} -1 } \int_1^{a_m^R}  \left[  a_m a^{\frac{\mu_m}{1-\mu}} ( a_m^{\gamma} p_u^*)^{1-\mu_m} \right]^{1-\sigma} N^{\frac{\mu_m}{1-\mu}} \cdot a_m^{\rho-1} da_m \\
\
&= \frac{\rho D^*}{ \overline{a}_m^{\rho} -1 } \left[  a^{\frac{\mu_m}{1-\mu}} ( p_u^*)^{1-\mu_m} \right]^{1-\sigma} N^{\frac{\mu_m}{1-\mu}} \int_1^{a_m^R} a_m^{\rho-1-(\sigma-1)[1+\gamma(1-\mu_m)]} da_m \\
\
&= \frac{\rho D^*}{ \overline{a}_m^{\rho} -1 } \left[  a^{\frac{\mu_m}{1-\mu}} ( p_u^*)^{1-\mu_m} \right]^{1-\sigma} N^{\frac{\mu_m}{1-\mu}} \frac{ (a_m^R)^{ \widetilde{\rho} } -1 }{ \widetilde{\rho} },
\end{align*}
where $\widetilde{\rho} \equiv \rho-(\sigma-1)[1+\gamma(1-\mu_m)]$ and $p_m^{1-\sigma}$ is given by Eq. (A3).

As in Appendix A1-3, we substitute these expressions into the zero-profit condition of host domestic firms, $pq=\sigma F$, to obtain
\begin{align*}
&pq = \sigma F, \\
\
&\to \frac{\sigma}{\sigma - \mu(\sigma-1)} \left[ \frac{E}{N} + \frac{\mu_m (\sigma-1)}{\sigma} \frac{K_f}{N} \int_1^{a_m^R} p_m(a_m) q_m(a_m)dG(a_m) \right] = \sigma F, \\
\
&\to \frac{\sigma}{\sigma(1-\mu) + \mu} \left[ \frac{\alpha L}{N}
+ \frac{\mu_m (\sigma-1)}{\sigma} \frac{K_f}{N}
\frac{\rho D^*}{ \overline{a}_m^{\rho} -1 } \left\{  a^{\frac{\mu_m}{1-\mu}} ( p_u^*)^{1-\mu_m} \right\}^{1-\sigma} N^{\frac{\mu_m}{1-\mu}} \frac{ (a_m^R)^{\widetilde{\rho}} -1 }{ \widetilde{\rho} } \right] = \sigma F, \\
\
&\to K_f[(a_m^R)^{\widetilde{\rho}} -1  ] = \widetilde{\Theta}_0 N^{ -\frac{\mu_m}{1-\mu} }(N - \alpha \overline{N}), \\
\
&\text{where} \ \ \widetilde{\rho} \equiv \rho-(\sigma-1)[1+\gamma(1-\mu_m)], \\
\
&\ \ \ \ \ \ \ \ \ \widetilde{\Theta}_0 \equiv \frac{\widetilde{\rho} (\overline{a}_m^{\rho} -1) \sigma F [\sigma (1-\mu) +\mu] \left[  a^{\frac{\mu_m}{1-\mu}} ( p_u^*)^{1-\mu_m} \right]^{1-\sigma}  }{ \rho \mu_m D^* (\sigma -1) }, 
\ \ \ \ \overline{N} \equiv \frac{L}{F[\sigma(1-\mu) +\mu]},
\end{align*}
noting that the upper bound of the number of host domestic firms, $\overline{N}$, is the same as that in the basic model.

Using Eq. (A5), we can rewrite the above equation as
\begin{align*}
&K_f \left[  \left\{ N_m ( \overline{a}_m^{\rho} -1 )/K_f + 1 \right\}^{\widetilde{\rho}/\rho}   -1  \right] = \widetilde{\Theta}_0 N^{ -\frac{\mu_m}{1-\mu} }(N - \alpha \overline{N}), \\
\
&\to N_m = \frac{K_f}{ \overline{a}_m^{\rho} -1 } \left[  \left\{ \widetilde{\Theta}  N^{ -\frac{\mu_m}{1-\mu} }(N - \alpha \overline{N})+ 1 \right\}^{\rho/\widetilde{\rho}} -1  \right], \tag{A7} \\
\
&\text{where} \ \ \widetilde{\Theta} \equiv \widetilde{\Theta}_0/K_f = \frac{\widetilde{\rho} (\overline{a}_m^{\rho} -1) \sigma F [\sigma (1-\mu) +\mu] \left[  a^{\frac{\mu_m}{1-\mu}} ( p_u^*)^{1-\mu_m} \right]^{1-\sigma}  }{ \rho \mu_m D^* K_f (\sigma -1) }.
\end{align*}
The $\Pi=0$-locus has an upward-sloping curve as we have assumed $\alpha > 1 - (1-\mu)/\mu_m$.
Fig. A1(a) draws the $\Pi=0$-locus with arrows indicating the direction of motion of host domestic firms.

\

\begin{center}
\includegraphics[scale=0.7]{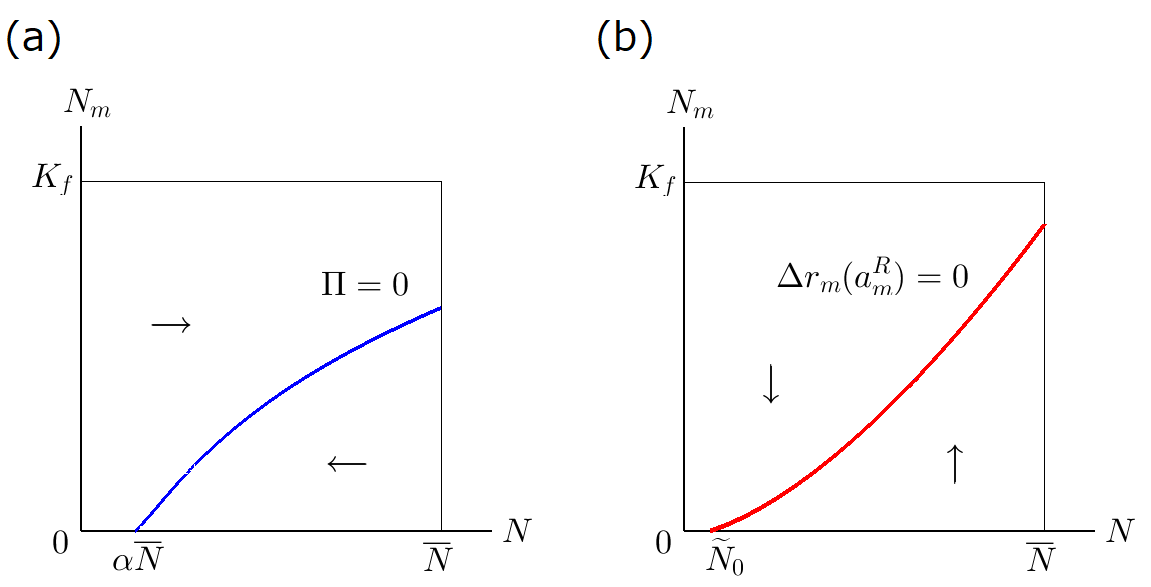} \\
Fig. A1. \ Equilibrium curves under heterogeneous multinationals
\end{center}

\

\subsubsection*{A3-3. \ \ Conditions for multiple equilibria}

We can tell from Figs. 6 and A1 that multiple equilibria occur if the two equilibrium curves intersect twice.
Sufficient conditions for this are as follows.
First, the $N$-intercept of the $\Pi=0$-curve is greater than or equal to that of the $\Delta r_m(a_m^R)=0$-curve, that is, $\alpha \overline{N} \ge \widetilde{N}_0$.
Second, the $\Pi=0$-curve is located above the $\Delta r_m(a_m^R)=0$-curve for some $N \in [\alpha \overline{N}, \overline{N}]$.
Finally, the $\Delta r_m(a_m^R)=0$-curve is located above the $\Pi=0$-curve at $N=\overline{N}$.

The first condition reduces to
\begin{align*}
&\alpha \overline{N} \equiv \frac{\alpha L}{F[\sigma(1-\mu) + \mu]} \ge \widetilde{N}_0 \equiv a^{\sigma-1} \left[  (p_u^*)^{-1} \right]^{(\sigma-1)(1-\mu)}, \\
\
&\to F \le \frac{\alpha L}{ \widetilde{N}_0 [\sigma(1-\mu) + \mu]} \equiv \widetilde{F}_b.
\end{align*}

The second condition requires that for some $N \in [\alpha \overline{N}, \overline{N}]$, the following holds:
\begin{align*}
&\frac{K_f}{ \overline{a}_m^{\rho} -1 } \left[  \left\{ \widetilde{\Theta}  N^{ -\frac{\mu_m}{1-\mu} }(N - \alpha \overline{N})+ 1 \right\}^{\rho/\widetilde{\rho}} -1  \right]
> \frac{K_f}{ \overline{a}_m^{\rho} -1 }
\left[ \left\{ a^{\frac{1}{1-\mu}} (p_u^*)^{-1} \right\}^{ - \frac{\rho \mu_m}{\gamma(1-\mu_m)} } N^{ \frac{\rho \mu_m}{\gamma(\sigma-1)(1-\mu)(1-\mu_m)} } - 1 \right], \\
\
&\to \left\{ \widetilde{\Theta}  N^{ -\frac{\mu_m}{1-\mu} }(N - \alpha \overline{N})+ 1 \right\}^{\rho/\widetilde{\rho}}
> \left\{ a^{\frac{1}{1-\mu}} (p_u^*)^{-1} \right\}^{ - \frac{\rho \mu_m}{\gamma(1-\mu_m)} } N^{ \frac{\rho \mu_m}{\gamma(\sigma-1)(1-\mu)(1-\mu_m)} }, \\
\
&\to \widetilde{\Theta}  N^{ -\frac{\mu_m}{1-\mu} }(N - \alpha \overline{N})+ 1
> \left\{ a^{\frac{1}{1-\mu}} (p_u^*)^{-1} \right\}^{ - \frac{\widetilde{\rho} \mu_m}{\gamma(1-\mu_m)} } N^{ \frac{\widetilde{\rho} \mu_m}{\gamma(\sigma-1)(1-\mu)(1-\mu_m)} }, \\
\
&\to \widetilde{\Theta}_1  N^{ -\frac{\mu_m}{1-\mu} } \left[ F N - \frac{\alpha L}{ \sigma(1-\mu) +\mu } \right]+ 1
> \left\{ a^{\frac{1}{1-\mu}} (p_u^*)^{-1} \right\}^{ - \frac{\widetilde{\rho} \mu_m}{\gamma(1-\mu_m)} } N^{ \frac{\widetilde{\rho} \mu_m}{\gamma(\sigma-1)(1-\mu)(1-\mu_m)} }, \\
\
&\to F > \max \{ F_1, 0\} \equiv \widetilde{F}_a, \\
\
&\text{where} \ \ F_1 \equiv \widetilde{\Theta}_1^{-1}  N^{\frac{\mu_m}{1-\mu} -1 } \left[ \left\{ a^{\frac{1}{1-\mu}}  (p_u^*)^{-1} \right\}^{ - \frac{\widetilde{\rho} \mu_m}{\gamma(1-\mu_m)} } N^{ \frac{\widetilde{\rho} \mu_m}{\gamma(\sigma-1)(1-\mu)(1-\mu_m)} } - 1 \right]
+ \frac{\alpha L}{ \sigma(1-\mu) +\mu },
\end{align*}
and $\widetilde{\Theta}_1 \equiv \widetilde{\Theta}/F$.

The third condition implies
\begin{align*}
&\frac{K_f}{ \overline{a}_m^{\rho} -1 }
\left[ \left\{ a^{\frac{1}{1-\mu}} (p_u^*)^{-1} \right\}^{ - \frac{\rho \mu_m}{\gamma(1-\mu_m)} } \overline{N}^{ \frac{\rho \mu_m}{\gamma(\sigma-1)(1-\mu)(1-\mu_m)} } - 1 \right]
> \frac{K_f}{ \overline{a}_m^{\rho} -1 } \left[  \left\{ \widetilde{\Theta} \overline{N}^{1 -\frac{\mu_m}{1-\mu} }(1 - \alpha)+ 1 \right\}^{\rho/\widetilde{\rho}} -1  \right], \\
\
&\to \left\{ a^{\frac{1}{1-\mu}} (p_u^*)^{-1} \right\}^{ - \frac{\rho \mu_m}{\gamma(1-\mu_m)} } \overline{N}^{ \frac{\rho \mu_m}{\gamma(\sigma-1)(1-\mu)(1-\mu_m)} }
> \left\{ \widetilde{\Theta} \overline{N}^{\frac{1-\mu-\mu_m}{1-\mu} }(1 - \alpha)+ 1 \right\}^{\rho/\widetilde{\rho}}, \\
\
&\to \left\{ a^{\frac{1}{1-\mu}}  (p_u^*)^{-1} \right\}^{ - \frac{\widetilde{\rho} \mu_m}{\gamma(1-\mu_m)} } \overline{N}^{ \frac{\widetilde{\rho} \mu_m}{\gamma(\sigma-1)(1-\mu)(1-\mu_m)} }
> K_f^{-1} \widetilde{\Theta}_0 \overline{N}^{\frac{1-\mu-\mu_m}{1-\mu} }(1 - \alpha)+ 1, \\
\
&\to K_f > \left[ a^{\frac{1}{1-\mu}}  (p_u^*)^{-1} \right]^{ \frac{\widetilde{\rho} \mu_m}{\gamma(1-\mu_m)} } \overline{N}^{ -\frac{\widetilde{\rho} \mu_m}{\gamma(\sigma-1)(1-\mu)(1-\mu_m)} }
\left[ \widetilde{\Theta}_0 \overline{N}^{\frac{1-\mu-\mu_m}{1-\mu} }(1 - \alpha)+ 1 \right] \equiv \widetilde{K}_f.
\end{align*}
In sum, the multiple equilibria occur if the fixed labor input takes an intermediate value such that $\widetilde{F}_a < F \le \widetilde{F}_b$ and the amount of foreign capital is so large that $K_f > \widetilde{K}_f$ holds.

\

\subsubsection*{A3-4. \ \ Proof of Proposition 3}

Since Proposition 3(i) is evident from Fig. 7 and the discussions in the main text, here we prove Proposition 3(ii). 
Assume that (a) $\alpha > 1 - (1-\mu)/\mu_m$; (b) $\widetilde{N}_0 \equiv [a(p_u^*)^{-(1-\mu)}]^{\sigma-1}>1$; (c) $K_f > \widetilde{K}_f$; (d) $F \in (\widetilde{F}_a, \widetilde{F}_b]$; and (e) $\widetilde{\rho} \equiv \rho - (\sigma-1)[1+\gamma(1-\mu_m)]$.
The first three assumptions are sort of regularity conditions.
Assumption (a) guarantees the upward slope of the $\Pi=0$-locus in the $(N, N_m)$ plane.
Assumptions (b) and (c) respectively ensure a finite value of expectation and a sufficient number of local suppliers at point $S_2$.
If (c) did not hold, the comparative statics with respect to the exponent of $N$ would yield meaningless results (see Eq. (A7)). 
Under the last two assumptions, (d) and (e), multiple equilibria arise. 

Consider an increase in the fixed labor input $F$ due to a natural disaster.
As in the basic model, this shock results in (A) a leftward shift of the vertical line $N= \overline{N}$; (B) an upward shift of the $\Pi=0$-curve; (C) no change in the $\Delta r_m(a_m^R)=0$-curve.
Fig. 7 illustrates these shifts of curves, where the $\Pi'=0$-curve and the $N=\overline{N}'$-line are the corresponding curves after the shock.

It can be seen from observations (B) and (C) that point $S_1$ in Fig. 7 is no longer a stable equilibrium if the upward shift of the $\Pi=0$-curve is so large that the new $\Pi'=0$-curve and the $\Delta r_m(a_m^R)=0$-curve intersect only once.
From Eq. (A7), the shift of the $\Pi=0$-curve is proportional to
\begin{align*}
0 < \frac{\partial N_m}{\partial F} &\propto 
\frac{\partial}{\partial F} \left[\widetilde{\Theta} N^{-\frac{\mu_m}{1-\mu}}(N -\alpha \overline{N}) \right] = \frac{\partial}{\partial F} \left[\widetilde{\Theta}_1 N^{-\frac{\mu_m}{1-\mu}} \left\{ F N -\frac{\alpha L}{\sigma(1-\mu) + \mu} \right\} \right] \\
\
&= \widetilde{\Theta}_1 N^{\frac{1-\mu-\mu_m}{1-\mu}} \\
\
&= \frac{\widetilde{\rho} (\overline{a}_m^{\rho} -1) \sigma [\sigma (1-\mu) +\mu] \left[  a^{\frac{\mu_m}{1-\mu}} (p_u^*)^{1-\mu_m} \right]^{1-\sigma}  }{ \rho \mu_m D^* K_f (\sigma -1) }.
\end{align*}
As $\partial N_m/\partial F$ is greater, point $S_1$ is less likely to be a stable equilibrium.
The magnitude of the upward shift $\partial N_m/\partial F$ depends on the cost share of local intermediate goods in multinational production, $\mu_m$, entering both the numerator and denominator of the term in the last line.
Under our assumption that $\widetilde{N}_0 > 1$, i.e., $\ln \widetilde{N}_0 > 0$, the numerator decreases with $\mu_m$:
\begin{align*}
\frac{\partial }{\partial \mu_m}  \left[  \left\{ a^{\frac{\mu_m}{1-\mu}} (p_u^*)^{1-\mu_m} \right\}^{1-\sigma} \right]
&= \left[ a^{\frac{\mu_m}{1-\mu}} (p_u^*)^{1-\mu_m} \right]^{1-\sigma} 
\ln \left[ a^{\frac{1}{1-\mu}} (p_u^*)^{-1} \right]^{1-\sigma} \\
\
&= (1-\mu)^{-1} \left[ a^{\frac{\mu_m}{1-\mu}} (p_u^*)^{1-\mu_m} \right]^{1-\sigma} 
\ln \left[ a (p_u^*)^{-(1-\mu)} \right]^{1-\sigma} \\
\
&= (1-\mu)^{-1} \left[ a^{\frac{\mu_m}{1-\mu}} (p_u^*)^{1-\mu_m} \right]^{1-\sigma} 
\ln \widetilde{N}_0^{-1} \\
\
&= -(1-\mu)^{-1} \left[ a^{\frac{\mu_m}{1-\mu}} (p_u^*)^{1-\mu_m} \right]^{1-\sigma} 
\ln \widetilde{N}_0 < 0.
\end{align*}
Since the denominator unambiguously increases with $\mu_m$, the whole term decreases with $\mu_m$:
\begin{align*}
\frac{\partial^2 N_m}{\partial F \partial \mu_m} &\propto 
\frac{\partial^2}{\partial F \partial \mu_m} \left[\widetilde{\Theta} N^{-\frac{\mu_m}{1-\mu}}(N -\alpha \overline{N}) \right] \\
\
&= \frac{\widetilde{\rho} (\overline{a}_m^{\rho} -1) \sigma [\sigma (1-\mu) +\mu] }{ \rho D^* K_f (\sigma -1) }
\frac{\partial}{\partial \mu_m} \left[  \mu_m^{-1} \left\{ a^{\frac{\mu_m}{1-\mu}} (p_u^*)^{1-\mu_m} \right\}^{1-\sigma} \right] \\
\
&= \frac{\widetilde{\rho} (\overline{a}_m^{\rho} -1) \sigma [\sigma (1-\mu) +\mu] }{ \rho D^* K_f (\sigma -1) }
\left[
\mu_m^{-1} \frac{\partial}{\partial \mu_m} \left\{ a^{\frac{\mu_m}{1-\mu}} (p_u^*)^{1-\mu_m} \right\}^{1-\sigma}
- \mu_m^{-2} \left\{ a^{\frac{\mu_m}{1-\mu}} (p_u^*)^{1-\mu_m} \right\}^{1-\sigma} \right] < 0.
\end{align*}
That is, the upward shift of the $\Pi=0$-curve is smaller as $\mu_m$ is higher.
This establishes Proposition 3(ii), stating that  multinationals emphasizing local sourcing show the resilience to natural disasters.

\

\subsection*{Appendix 4. \ \ Reconstruction from disasters}

Supposing that a natural disaster strikes at an initial time $s=0$, the fixed labor input for host domestic firms at time $s \ge 0$ is specified as $F(s) = F e^{\delta (T-s)}$.
The fixed input returns to the pre-shock level $F$ after time $T$.
At point $S_1$, capital return of becoming a multinational at time $s \ (\le T)$, denoted by $r_m(s)$, is given by
\begin{align*}
r_m(s) &= (D^*/\sigma) \left[ a_m a^{\frac{\mu_m}{1-\mu}}  (\tau p_u^*)^{1-\mu_m} \right]^{1-\sigma} \overline{N}(s)^{\frac{\mu_m}{1-\mu}} \\
\
&= (D^*/\sigma) \left[ a_m a^{\frac{\mu_m}{1-\mu}}  (\tau p_u^*)^{1-\mu_m} \right]^{1-\sigma} 
\left[ \frac{L}{F(s)\{ \sigma(1-\mu) + \mu\}}
\right]^{\frac{\mu_m}{1-\mu}} \\
\
&= (D^*/\sigma) \left[ a_m a^{\frac{\mu_m}{1-\mu}}  (\tau p_u^*)^{1-\mu_m} \right]^{1-\sigma} 
\left[ \frac{L}{F e^{\delta (T-s)}\{ \sigma(1-\mu) + \mu\}}
\right]^{\frac{\mu_m}{1-\mu}} \\
\
&= (D^*/\sigma) \left[ a_m a^{\frac{\mu_m}{1-\mu}}  (\tau p_u^*)^{1-\mu_m} \right]^{1-\sigma} \overline{N}^{\frac{\mu_m}{1-\mu}}
e^{-\frac{\delta \mu_m(T-s)}{1-\mu}} \\
\
&\equiv r_m e^{-\frac{\delta \mu_m(T-s)}{1-\mu}} \ \ \ \ \text{for} \ s < T,
\end{align*}
where we slightly abuse the notation of $r_m$
and note that the number of host domestic firms is  $N=\overline{N}$ at $S_1$.
We can check $r_m \ge r_f = (a_m p_u^*)^{1-\sigma}D^*/\sigma$ if $\overline{N} \ge N_0$ (defined in Eq. (13)), or equivalently $F \le F_b \equiv L/N_0[\sigma(1-\mu)+\mu]$.

Letting $t \ (\le T)$ be the time at which foreign capital moves from the foreign to the host country, its lifetime return is given by
\begin{align*}
v(t) &\equiv \int_0^t e^{-\theta s} r_f ds + \int_t^{\infty} e^{-\theta s} r'_m(s)ds \\
\
&= \int_0^t e^{-\theta s} r_f ds + \int_t^{T} e^{-\frac{\theta \delta \mu_m(T-s)}{1-\mu}} r_mds
+ \int_T^{\infty} e^{-\theta s} r_m ds
, \ \ \ \ \text{for} \ t \le T,
\end{align*}
where $\theta>0$ is the discount rate.
The optimal timing of entering the host country is derived from the following FOC:
\begin{align*}
&v'(t) = e^{-\theta t} r_f - e^{-\frac{\theta \delta \mu_m(T-t)}{1-\mu}} r_m = 0, \\
\
&\to -\theta t + \ln r_f = -\frac{\theta \delta \mu_m(T-t)}{1-\mu} + \ln r_m, \\
\
&\to \frac{\theta (1 -\mu +\delta \mu_m)t}{1-\mu} = \frac{\theta \delta \mu_m T}{1-\mu} 
+ \ln \left( \frac{r_f}{r_m} \right), \\
\
&\to t =\begin{cases} 0 &\text{if} \ \ T \le \widehat{T} \equiv \dfrac{1-\mu}{\theta \delta \mu_m} \ln \left(   \dfrac{r_m}{r_f} \right) \\
\dfrac{1}{1 -\mu +\delta \mu_m} \left[ \delta T - \dfrac{1-\mu}{\theta} \ln \left( \dfrac{r_m}{r_f} \right) \right] \equiv \widehat{t} \ (> 0) &\text{if} \ \ T > \widehat{T}
\end{cases}, \tag{A8}
\end{align*}
noting that $r_m > r_f$, or equivalently $\ln (r_m/r_f)>0$ holds at $S_1$.
We can immediately see from Eq. (A8) that the interior optimal timing increases with the recovery time, i.e., $\partial \widehat{t}/\partial T > 0$.
It is indeed smaller than the recovery time $T$:
\begin{align*}
T - \widehat{t} = \frac{1-\mu}{1-\mu+\delta \mu_m} \left[ T + \frac{1}{\theta} \ln \left( \frac{r_m}{r_f}\right) \right] > 0.
\end{align*}

The second-order condition (SOC) is
\begin{align*}
&v''(t) =-\theta e^{-\theta t} r_f + \frac{\theta \delta \mu_m}{1-\mu} e^{-\frac{\theta \delta \mu_m(T-t)}{1-\mu}} r_m < 0, \\
\
&\to \frac{\delta \mu_m}{1-\mu} \cdot \exp\left( \theta \left[ t - \frac{\delta \mu_m(T-t)}{1-\mu} \right] \right) < \frac{r_f}{r_m}, \\
\
&\to \ln \left( \frac{\delta \mu_m}{1-\mu} \right) + \theta \left[ t + \frac{\delta \mu_m(t-T)}{1-\mu} \right] < \ln \left( \frac{r_f}{r_m} \right).
\end{align*}
If the SOC does not hold, the objective function $v(t)$ exhibits a convex one and foreign capital never moves to the host country, i.e., $t= \infty$.
The inequality in the last line always holds if the following sufficient condition holds:
\begin{align*}
&\ln \left( \frac{\delta \mu_m}{1-\mu} \right) + \theta \left[ t + \frac{\delta \mu_m(t-T)}{1-\mu} \right]
\le \ln \left( \frac{\delta \mu_m}{1-\mu} \right) + \theta \left[ T + \frac{\delta \mu_m(T-T)}{1-\mu} \right] < \ln \left( \frac{r_f}{r_m} \right), \\
\
&\to \ln \left( \frac{\delta \mu_m}{1-\mu} \right) + \theta T < \ln \left( \frac{r_f}{r_m} \right), \\
\
&\to T < \overline{T} \equiv \frac{1}{\theta} \ln \left( \frac{1-\mu}{\delta \mu_m} \cdot \frac{r_f}{r_m} \right),
\end{align*}
where $\overline{T}$ is defined over a parameter range that ensures $\overline{T}>0$.
If $T$ is sufficiently higher than $\overline{T}$, to the contrary, the SOC is not satisfied and the time of reentering never comes, i.e., $t=\infty$, which establishes Proposition 4(i). 

\pagebreak

Assuming the existence of interior solutions (a parameter range that satisfies $\widehat{t}<\overline{T}$), positive trade costs ($\tau>1$), and $T = \widehat{T} < \overline{T}$,
we can check that the optimal timing $t=\widehat{t}$ decreases with $\mu_m$:
\begin{align*} \frac{\partial \widehat{t}}{\partial \mu _m}&= \frac{1}{(1-\mu +\delta \mu _m)^2} \left[ \left\{\delta \widehat{T} -\frac{1-\mu }{\theta } \frac{\partial \ln (r_m/r_f)}{\partial \mu _m}\right\} \cdot (1-\mu +\delta \mu _m) -\delta \left\{ \delta \mu_m \widehat{T} -\frac{1-\mu }{\theta } \ln \left( \frac{r_m}{r_f} \right) \right\} \right] \\&= \frac{1}{(1-\mu +\delta \mu _m)^2} \cdot \frac{(1-\mu )(1-\mu +\delta \mu _m)}{\theta \mu _m} \ln \tau ^{1-\sigma }  < 0. \end{align*}
Because the derivative increases with $T$, it still becomes negative if $T$ is greater than but close to $\widehat{T}$.
From the first to the second line, the following relations were used:
\begin{align*}
&\frac{r_m}{r_f} = \left[ \left\{ a^{\frac{1}{1-\mu}} (\tau p_u^*)^{-1}  \right\}^{1-\sigma} \overline{N}^{\frac{1}{1-\mu}} \right]^{\mu_m} \tau^{1-\sigma} > 1, \\
\
&\frac{\partial (r_m/r_f)}{\partial \mu_m}
= r_m \ln \left[ \left\{ a^{\frac{1}{1-\mu}} (\tau p_u^*)^{-1} \right\}^{1-\sigma} \overline{N}^{\frac{1}{1-\mu}} \right] \\
\
&\ \ \ \ \ \ \ \ \ \ \ \ = \frac{1}{\mu_m} \frac{r_m}{r_f} \left[ \ln \left( \frac{r_m}{r_f} \right) - \ln \tau^{1-\sigma} \right] > 0. 
\end{align*}
Similarly, it can be checked that $\partial \widehat{t}/\partial \tau < 0$.
These establish Proposition 4(ii).

\

\subsection*{Appendix 5. \ \ Host country's market}

We here modify the basic model to allow multinationals to serve the final goods market in the host country.
For simplicity, we assume that host domestic firms do not produce final goods.
Fig. A2 depicts the structure of the modified setting.

\

\begin{center}
\includegraphics[scale=1.0]{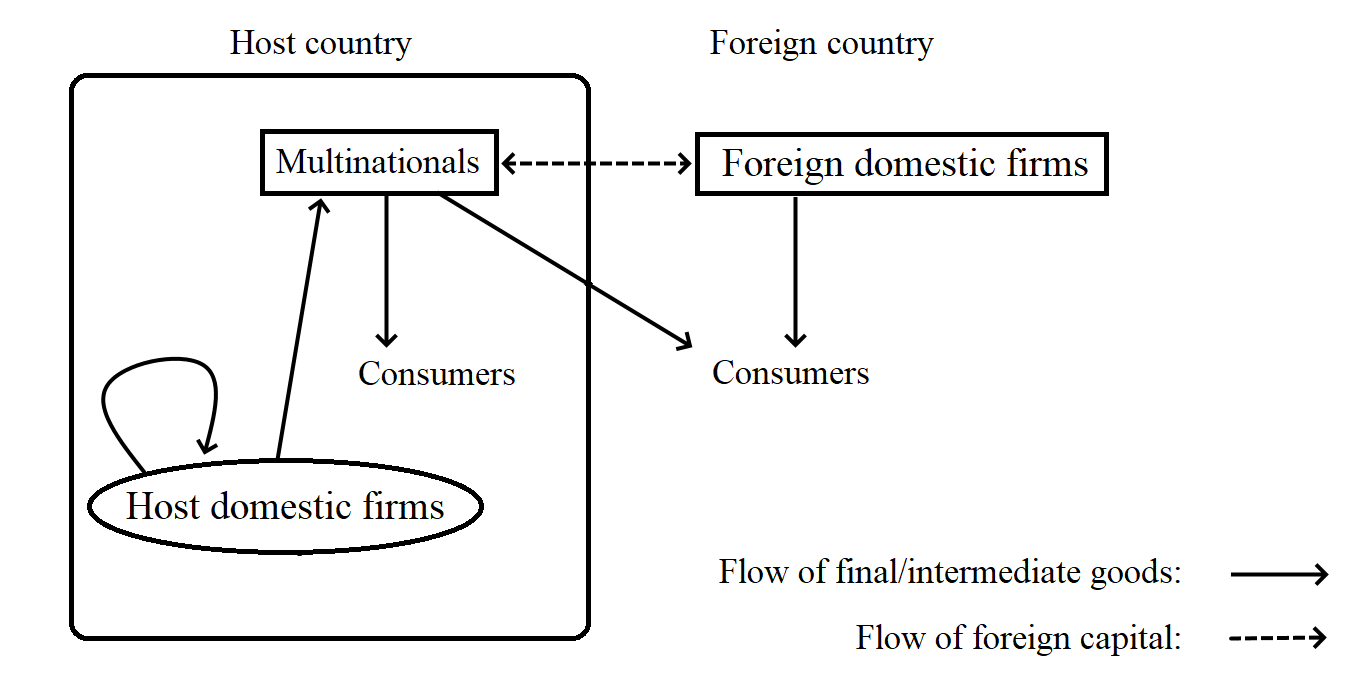} \\
Fig. A2. \ Model with the host country's market
\end{center}

\

Since there are two price indices in the host country, we change notations and denote the price index of intermediate goods by $P_u$ and that of final goods by $P$.
The demand functions for a typical variety produced by multinationals and foreign domestic firms are respectively 
\begin{align*}
&q_m = \left( \frac{p_m}{P} \right)^{-\sigma} \frac{E}{P}
+ \left( \frac{p_m}{P^*} \right)^{-\sigma} \frac{E^*}{P^*}, \\
\
&q_f = \left( \frac{p_f}{P^*} \right)^{-\sigma} \frac{E^*}{P^*},
\end{align*}
where the price indices are defined as
\begin{align*}
&P = \left[ N_m p_m^{1-\sigma} + (K_f-N_m) p_f^{1-\sigma} \right]^{\frac{1}{1-\sigma}}, \\
\
&p_m = a_m P_u^{\mu_m} (\tau p_u^*)^{1-\mu_m}, \\
\
&P_u = N^{\frac{1}{1-\sigma}} p.
\end{align*}
Other variables are defined as in the main text.
The supply of intermediate goods by host domestic firms is equal to the sum of demand by themselves and by  multinationals:
\begin{align*}
q &= N \mu p^{-\sigma} P_u^{\sigma +\mu -1} w^{1-\mu}  \widetilde{a} q + N_m \mu_m p^{-\sigma} P_u^{\sigma +\mu_m -1} (\tau p_u^*)^{1-\mu_m} \widetilde{a}_m q_m \\
\
&= p^{-\sigma} \left[ N \mu P_u^{\sigma +\mu -1} \widetilde{a} q + N_m \mu_m P_u^{\sigma +\mu_m -1} (\tau p_u^*)^{1-\mu_m} \widetilde{a}_m q_m \right].
\end{align*} 
which is different from Eq. (11) in that the price index of intermediate goods is given by $P_u$ and that there are no consumer demand for locally-produced final goods.

As in Appendix A1-3, we can derive sales of a typical intermediate-good variety as
\begin{align*}
pq &= \frac{\sigma}{\sigma - \mu(\sigma-1)} \frac{\mu_m (\sigma-1)}{\sigma} \frac{N_m}{N} p_m q_m \\
\
&= \frac{\sigma}{\sigma - \mu(\sigma-1)} \frac{\mu_m (\sigma-1)}{\sigma} \frac{N_m}{N} p_m^{1-\sigma} (EP^{\sigma-1} +D^*),
\end{align*}
where $D^* \equiv E^* (P^*)^{\sigma-1}$.

Using this, we can derive the combinations of $N$ and $N_m$ that make excess profits zero:
\begin{align*}
&\Pi = pq - P_u^{\mu} \widetilde{a} q - F = 0, \\
\ 
&\to pq = \sigma F, \\
\
&\to N_m = \Theta N^{-\frac{\mu_m}{1-\mu}}  \left[ N  -\alpha \mu_m(\sigma-1)\overline{N}/\sigma \right], \\
\
&\text{where} \ \ \Theta \equiv \frac{ \sigma F  [\sigma(1-\mu)+\mu] \left[ a_m a^{\frac{\mu_m}{1-\mu}}  (\tau p_u^*)^{1-\mu_m} \right]^{\sigma-1} }{\mu_m D^*(\sigma-1) }, \ \ \ \
\overline{N} \equiv \frac{L}{F[\sigma(1-\mu)+\mu]}.
\end{align*}
Assuming the gradual entry-and-exit process, the number of host domestic firms increases if the excess profit is positive, i.e., $\Pi = pq -\sigma F >0$, or equivalently $N_m > \Theta N^{-\frac{\mu_m}{1-\mu}}  \left[ N  -\alpha \mu_m(\sigma-1)\overline{N}/\sigma \right]$, and it decreases otherwise.

Turning to multinationals and foreign domestic firms, their rental rate of capital is respectively
\begin{align*}
&r_m = p_m^{1-\sigma} (EP^{\sigma-1} +D^*)/\sigma, \\
\
&r_f = p_f^{1-\sigma} D^*/\sigma.
\end{align*}
Taking the difference of the two gives
\begin{align*}
\Delta r_m &\equiv r_m -r_f \\
\
&= D^* (a_m p_u^*)^{1- \sigma} \left[  \left\{ a^{\frac{\mu_m}{1-\mu}} \tau^{1-\mu_m} (p_u^*)^{-\mu_m} \right\}^{1-\sigma} N^{\frac{\mu_m}{1-\mu}} -  1 \right]/\sigma
+ \alpha L/(\sigma N_m),
\end{align*}
noting that the big square bracket term is positive if $N>N_0$, where $N_0$ is defined in Eq. (13). 
Assuming the gradual relocation process, the number of multinationals in the host country always increases if $N>N_0$.
Supposing $N \le N_0$, the number of multinationals increases if
\begin{align*}
&\Delta r_m > 0, \\
\
&\to N_m < \frac{\alpha L}{D^*} \frac{(a_m p_u^*)^{\sigma-1} }{ 1 - \left[ a^{\frac{\mu_m}{1-\mu}} \tau^{1-\mu_m} (p_u^*)^{-\mu_m} \right]^{1-\sigma} N^{\frac{\mu_m}{1-\mu}} },
\end{align*}
and it decreases otherwise. 

To make the modified model comparable with the basic model, we further assume (a) $K_f > N^{\frac{1-\mu-\mu_m}{1-\mu}}[1 -\alpha \mu_m(\sigma-1)/\sigma]$ to ensure the $\Pi=0$-locus never touches the $N_m=K_f$-line; and (b) $\mu+\mu_m<1$ to ensure the $\Pi=0$-locus has an upward slope.\footnote{Assumption (b) is a sufficient condition for $\partial N_m/\partial N = \Theta N^{-1-\frac{\mu_m}{1-\mu}} [\alpha \mu_m^2(\sigma-1)\overline{N}/\sigma+(1-\mu-\mu_m)N]>0$.}
Under these two assumptions, Figs. A3(a) and A3(b) draw the $\Pi=0$-locus and the $\Delta r_m=0$-locus respectively, with arrows indicating the direction of motion of host domestic firms and multinationals.

\

\begin{center}
\includegraphics[scale=0.9]{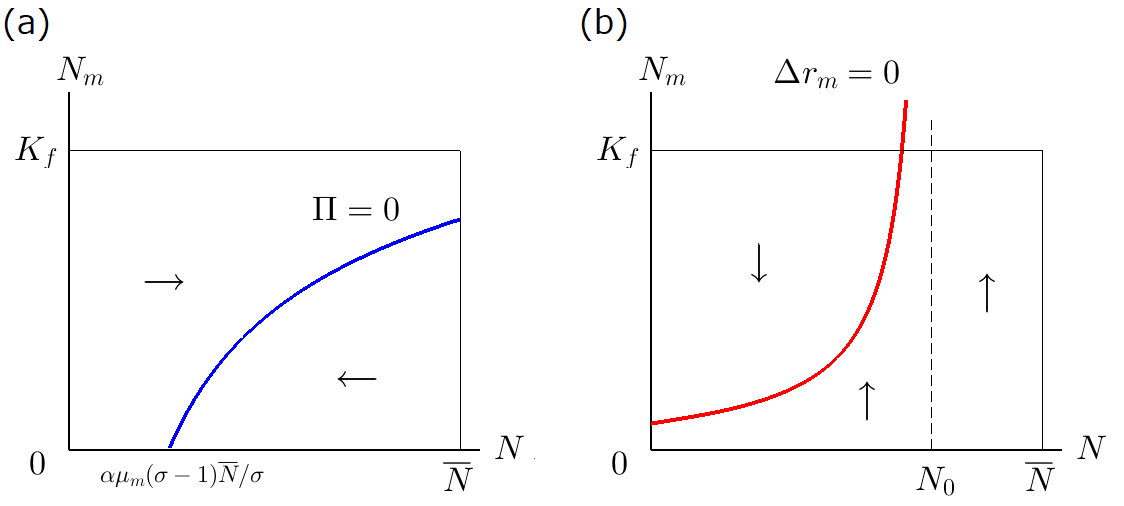} \\
Fig. A3. \ Equilibrium curves when multinationals serve the host market
\end{center}

\

There are two types of equilibrium configurations, one shown in Fig. A4(a) and the other in Fig. A4(b), except for a non-generic case where two equilibrium curves are tangent to each other.
Unlike Fig. 3 in the main text, we see a small but positive number of multinationals at point $S_1$ in Fig. A4(a).
Noting that the intercept of $\Delta r_m=0$-locus is always positive ($\alpha L(a_m p_u^*)^{\sigma-1}/D^*>0$), the $\Delta r_m=0$-locus intersects with the $\Pi=0$-locus at a positive $N$.
If the two curves do not intersect, only $S_2$, where multinationals and host domestic firms coexist, is the unique stable equilibrium as shown in Fig. A4(b).
In both cases, in contrast to the basic model, we see at least some units of foreign capital entering the host country in equilibrium.

The intuition is simple: introducing a final good market in the host country for multinationals makes becoming a multinational more profitable than becoming a foreign local firm.
We can also check that an increase in the host's market size, $\alpha L$, moves the $\Pi=0$-locus right and shifts up the $\Delta r_m=0$-locus.
This change shifts the two curves away from each other and thus point $S_2$ is more likely to be the unique stable equilibrium.
Put differently, as the host market is larger, foreign capital is more likely to enter the host country.

\

\begin{center}
\includegraphics[scale=0.6]{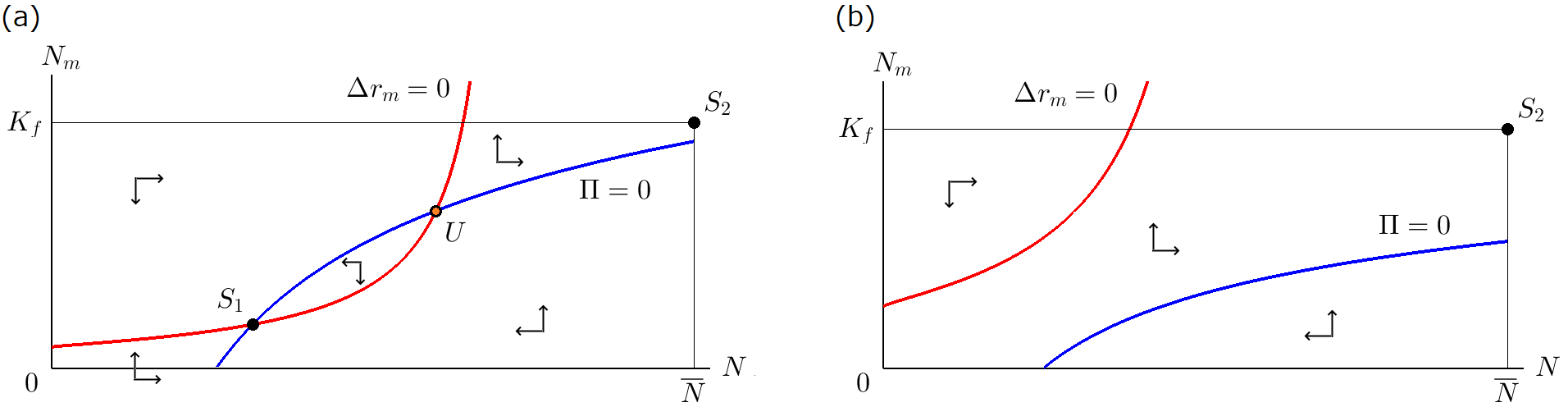} \\
Fig. A4. \ Equilibrium configurations when multinationals serve the host market
\end{center}

\

The findings are summarized as follows:

\

\noindent
{\bf Proposition A1 (Host country's market).} \ \ {\it Assume that multinationals serve the host market as well as the foreign market, while host domestic firms do not.
Assume also (a) $K_f > N^{\frac{1-\mu-\mu_m}{1-\mu}}[1 -\alpha \mu_m(\sigma-1)/\sigma]$; and (b) $\mu+\mu_m<1$.
Then, there may be two equilibrium configurations: $S_1$, where all units of foreign capital enter the host country as multinationals, and $S_2$, where some units of foreign capital do so.}

\

\subsection*{Appendix 6. \ \ Endogenous sourcing patterns}

We here allow foreign capital to choose the cost share of local intermediate goods, $\mu_m$, when locating in the host country.
There are two types of multinationals, one with high $\mu_m^H$ called a $H$-multinational and the other with low $\mu_m^L (< \mu_m^H)$ called a $L$-multinational.
To build a tighter relationship with local suppliers when starting operation, the $H$-multinational is likely to incur a greater fixed cost than the $L$-multinational.
Letting $F^j$ be the fixed capital input for the $j \in \{H, L\}$-multinational, this means $F^H>F^L>1$.

As in Appendix 1, we derive the rental rate of capital for $j \in \{ H, L\}$-multinational as
\begin{align*}
&r_m^j = p_m^j q_m^j/(\sigma F^j) = (p_m^j)^{1-\sigma} D^*/(\sigma F^j), \ \ \ \ j \in \{ H, L\}, \\
\
&\text{where} \ \ (p_m^j)^{1-\sigma} = \left[ a_m P^{\mu_m^j} (\tau p_u^*)^{1-\mu_m^j} \right]^{1-\sigma} 
= \left[ a_m a^{\frac{\mu_m^j}{1-\mu}}  (\tau p_u^*)^{1-\mu_m^j} \right]^{1-\sigma} N^{\frac{\mu_m^j}{1-\mu}},
\end{align*}
where the superscript $j$ represents the type of multinational.
The rental rate of capital for foreign domestic firms remains unchanged and is given by $r_f=p_f^{1-\sigma}D^*/\sigma = (a_m p_u^*)^{1-\sigma}D^*/\sigma$.
Comparing the three rental rates, we see 
\begin{align*}
r_f > r_m^L > r_m^H \ \ \ \ &\text{for} \ N \in [0, N_0), \\
\
r_m^L \ge \max\{ r_m^H, r_f \}  \ \ \ \ &\text{for} \ N \in [N_0, N_1], \\
\
r_m^H > r_m^L > r_f  \ \ \ \ &\text{for} \ N \in (N_1, \infty),
\end{align*}
where $N_0$ and $N_1$ are respectively defined as
\begin{align*}
&N_0 \equiv a^{\sigma-1} \left[ \tau^{\frac{1-\mu_m^L}{\mu_m^L}} (p_u^*)^{-1} \right]^{(\sigma-1)(1-\mu)} (F^L)^{\frac{1-\mu}{\mu_m^L}}, \ \ \text{at which $r_m^L=r_f$ holds}, \\
\
&N_1 \equiv a^{\sigma-1} ( \tau p_u^* )^{(1-\sigma)(1-\mu)} (F^H/F^L)^{\frac{1-\mu}{\mu_m^H-\mu_m^L}}, \ \ \text{at which $r_m^H=r_m^L$ holds}.
\end{align*}
We slightly abuse the notation of $N_0$, which is different from $N_0$ defined in Eq. (13) in the main text.
Our assumptions that $\mu_m^H>\mu_m^L$ and $F^H>F^L >1$ ensure the above inequalities on the rental rates and also the following inequality:
\begin{align*}
&N_1 > N_2 > N_0, \\
\
&\text{where} \ \ N_2 \equiv a^{\sigma-1} \left[ \tau^{\frac{1-\mu_m^H}{\mu_m^H}} (p_u^*)^{-1} \right]^{(\sigma-1)(1-\mu)} (F^H)^{\frac{1-\mu}{\mu_m^H}}, \ \ \text{at which $r_m^H=r_f$ holds}.
\end{align*}

Under the gradual relocation process, the ranking of capital return implies the following movement.
Letting $N_m^j$ be the number of type-$j \in \{ H, L\}$ multinationals and $N_f$ be the number of foreign domestic firms, as time goes by, $N_f$ increases if $N \in [0, N_0)$; $N_m^L$ increases if $N \in [N_0, N_1)$; $N_m^H$ increases if $N \in [N_1, \infty)$.
Conditional on entering the host country, multinationals choose a higher degree of local sourcing by incurring higher fixed costs $F^H$ if there are many local suppliers.

Turning to the host domestic firm, supposing that $N_1 < \overline{N}$, the zero-profit condition implies
\begin{align*}
&\Pi = pq -C(q) = 0, \\
\
&\to N_m = \begin{cases}
\Theta^L N^{-\frac{\mu_m^L}{1-\mu}}  \left( N  -\alpha \overline{N} \right) &\text{if} \ \ N \in [0, N_1] \\
\Theta^H N^{-\frac{\mu_m^H}{1-\mu}}  \left( N  -\alpha \overline{N} \right) &\text{if} \ \ N \in (N_1, \overline{N}] 
\end{cases}, \\
\
&\text{where} \ \ \Theta^j \equiv \frac{ \sigma F  [\sigma(1-\mu)+\mu] \left[ a_m a^{\frac{\mu_m^j}{1-\mu}}  (\tau p_u^*)^{1-\mu_m^j} \right]^{\sigma-1} }{\mu_m^j D^*(\sigma-1) }, \ \
\overline{N} \equiv \frac{L}{F[\sigma(1-\mu)+\mu]}, \ \ \ \ \text{for} \ j \in \{ H, L\},
\end{align*}
and where we can check that $\Theta^L N^{-\frac{\mu_m^L}{1-\mu}} > \Theta^H N^{-\frac{\mu_m^H}{1-\mu}}$.
The $\Pi=0$-locus is discontinuous at $N_1$, where conditional on entering the host country foreign capital is indifferent between becoming the $H$-multinational and the $L$-multinational; 
It chooses to become the $L$-multinational if $N \le N_1$ (or equivalently $r_m^L \le r_m^H$) and it chooses to become the $H$-multinational otherwise.

Under the gradual entry-and-exit process, $N$ increases over time if $N_m > \Theta^j N^{-\frac{\mu_m^j}{1-\mu}} (N - \alpha \overline{N})$ and decreases otherwise.
A typical equilibrium configuration is shown in Fig. A5.
At $S_1$, all foreign capital in the host country chooses to become the $H$-multinational.

\

\begin{center}
\includegraphics[scale=0.9]{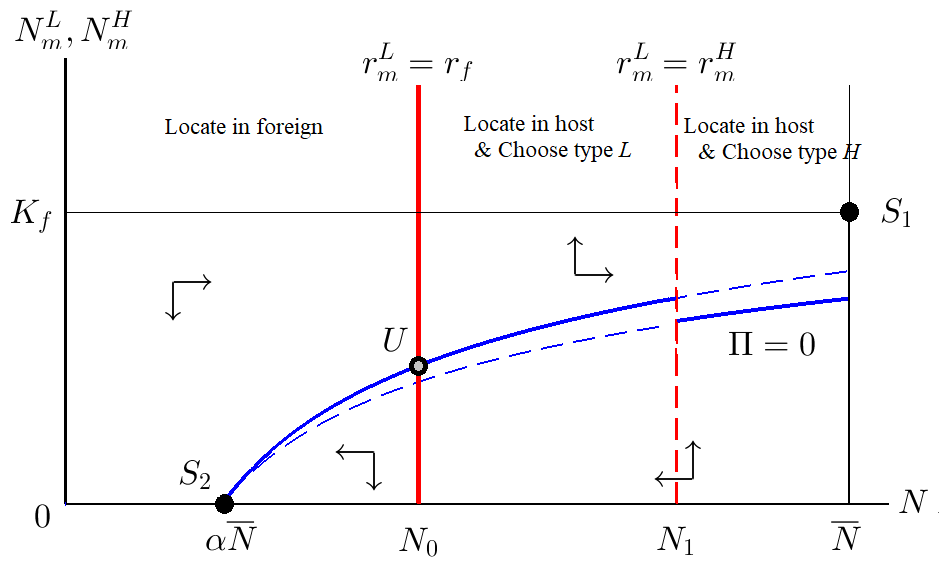} \\
Fig. A5. \ Multiple equilibria under endogenous sourcing patterns
\end{center}

\

Suppose that the host economy is initially at point $S_1$ and is hit by a natural disaster, raising the fixed labor input for host domestic firms  from $F$ to $F'$, where $F' \in (F_b, F_c)$, $F_b \equiv L/N_0[\sigma(1-\mu)+\mu]$ and $F_c \equiv L/N_1[\sigma(1-\mu)+\mu]$.
As a result, the equilibrium changes from $S_1$ to $S_1'$: all foreign capital switches from the $H$-multinational to the $L$-multinational, as illustrated in Fig. A5.
The increased cost of sourcing local intermediates due to the disaster is not severe enough for foreign capital to leave the host country, but severe enough for it to reduce the degree of local sourcing.
The damaged local industry decreases the profitability of local sourcing and is thus unable to help foreign capital choose the type $H$, which needs higher fixed capital requirement $F^H$.

\

\begin{center}
\includegraphics[scale=0.9]{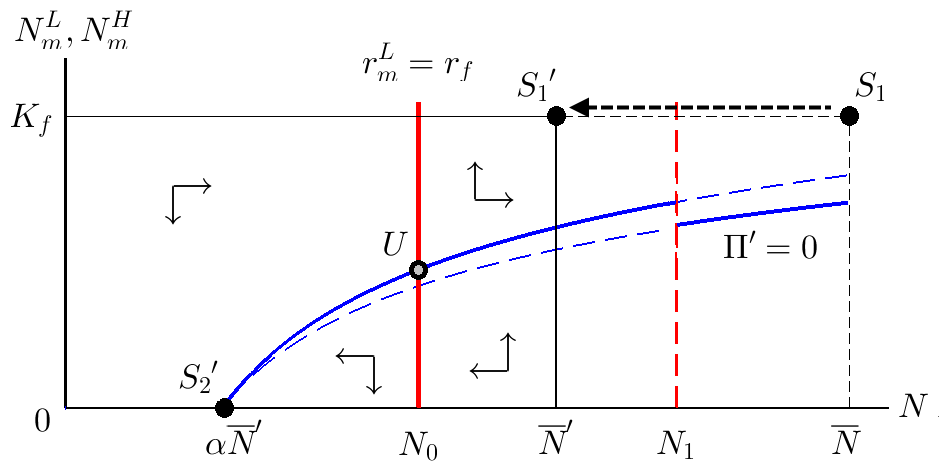} \\
Fig. A6. \ Disaster impact under endogenous sourcing patterns
\end{center}

\

In sum, assuming (a)$\alpha > 1-(1-\mu)/\mu_m^H$; (b)$K_f > \Theta^L \overline{N}^{\frac{1-\mu-\mu_m^L}{1-\mu}}(1-\alpha)$; and (c)$N_1 < \overline{N}$, we obtain the following proposition.

\

\noindent
{\bf Proposition A2 (Disaster impact under endogenous sourcing patterns).} \ \ {\it The basic model is modified in a way such that conditional on entering the host country foreign capital chooses to become an either of the two types of multinationals, $H$ or $L$.
The $H$-multinational has a higher cost share in local intermediate goods and a higher fixed capital input than the $L$-multinational, i.e., $\mu_m^H>\mu_m^L$ and $F^H>F^L>1$.

Consider the situation where the host economy is initially at point $S_1$ and all foreign capital in the host country chooses to become the $H$-multinational.
Supposing that a natural disaster hits there and the fixed labor input for host domestic firms increases from $F$ to $F' \in (F_b, F_c)$, where $F_b \equiv L/N_0[\sigma(1-\mu)+\mu]$ and $F_c \equiv L/N_1[\sigma(1-\mu)+\mu]$,
all the $H$-multinationals switch to the $L$-multinational and decrease the degree of local sourcing.}

\

\subsection*{Appendix 7. \ \ Disaster risk and the timing of leaving}

We here examine the effect of disaster risk on the location choice of foreign capital by introducing uncertainty on the timing of shock.
As in the analysis of the timing of reentering in Section 5.2, foreign capital makes dynamic location decisions so as to maximize its lifetime return.
Let us consider the situation where all foreign capital is initially located in the host country, corresponding to equilibrium $S_1$, and a shock hits there at time $s=T>0$.
The shock raises the fixed labor input for host domestic firms from $F$ to $F'$, where we assume $F'<F_b \equiv L/N_0[\sigma(1-\mu)+\mu]$ and $N_0$ is defined in Eq. (13).
The flow returns of multinationals and foreign domestic firms are
\begin{align*}
&r_m = (D^*/\sigma) \left[ a_m a^{\frac{\mu_m}{1-\mu}}  (\tau p_u^*)^{1-\mu_m} \right]^{1-\sigma} \overline{N}^{\frac{\mu_m}{1-\mu}}, \ \ \ \ \text{for} \ s \in [0, T) \\
\
&r_m' = (D^*/\sigma) \left[ a_m a^{\frac{\mu_m}{1-\mu}}  (\tau p_u^*)^{1-\mu_m} \right]^{1-\sigma} \left( \overline{N}' \right)^{\frac{\mu_m}{1-\mu}}, \ \ \ \ \text{for} \ s \in [T, \infty) \\
\
&r_f = p_f^{1-\sigma} D^*/\sigma, \ \ \ \ \text{for} \ s \in [0, \infty), \\
\
&\text{where} \ \ \overline{N} \equiv \frac{L}{F[\sigma(1-\mu) +\mu]} > \frac{L}{F'[\sigma(1-\mu) +\mu]} \equiv \overline{N}'.
\end{align*}
Under $F < F_b$, we have $r_m > r_m' > r_f$.
The shock is so small that staying in the disaster-hit host country is still more profitable than in the foreign country.

\

{\it If the timing of shock is known.} \ \ Suppose that multinationals know the exact timing of shock, which occurs at time $s=T$.
 Let $t$ be the time at which the multinational leaves the host country.
The lifetime return of foreign capital calculated at time $s=0$ is
\begin{align*}
&v(t) = \underbrace{\int_0^T e^{-\theta s} r_m ds + \int_T^t e^{-\theta s} r_m' ds}_{\text{Locate in host}} 
+ \underbrace{\int_t^{\infty} e^{-\theta s} r_f ds}_{\text{Locate in foreign}}.
\end{align*}
It can be easily seen that the optimal behavior is to stay in the host country even after the shock occurs, i.e., $t=\infty$.
This is because
\begin{align*}
v(t) &= \int_0^T e^{-\theta s} r_m ds + \int_T^t e^{-\theta s} r_m' ds + \int_{t}^{\infty} e^{-\theta s} r_f ds \\
\
&< \int_0^T e^{-\theta s} r_m ds + \int_T^{\infty} e^{-\theta s} r_m' ds = v(t=\infty).
\end{align*}
Without uncertainty on the timing of shock, multinationals never leave the host country.

\

{\it If the timing of shock is unknown.} \ \ 
Suppose that multinationals do not know the timing of shock $T$ and only know it follows a exponential distribution with a cumulative density function such that $G(T) = 1- e^{-\lambda T}$.
Note that $\lambda$ is the average arrival rate of shock per unit of time and a higher $\lambda$ means a higher frequency of shocks.
Suppose also that after the shock hits at time $T$ multinationals expect huge subsequent shocks and leave the host country before the next one comes.

The expected lifetime return of foreign capital is then
\begin{align*}
\E[v(t)] &= \underbrace{\int_0^{t} \left[ \int_0^T e^{-(\theta +\lambda)} r_mds + \int_T^t e^{-(\theta +\lambda) s}r_m'ds \right] e^{-\lambda T}dT}_{\text{Locate in host}}
+ \underbrace{\int_t^{\infty} e^{-\theta s} r_f ds}_{\text{Locate in foreign}} \\
\
&= \frac{\lambda (r_m-r_m')}{(\theta +\lambda)(\theta +2\lambda)} \left[1 - e^{-(\theta +2\lambda)t} \right]
+ \frac{1}{\theta+\lambda} \left( 1-e^{-\lambda t}\right) \left[ r_m - e^{-(\theta+\lambda)t} r_m'\right]
+ \frac{r_f}{\theta} e^{-\theta t},
\end{align*}
noting for example that the probability of the first shock not hitting until time $s < T$ is $1-G(s)= e^{-\lambda s}$ and thus the expected return of $r_m$ at time $s < T$ is $e^{-\theta s}r_m [1-G(s)] = e^{-(\theta +\lambda) s} r_m$.

Under a sufficiently small discount rate $\theta \simeq 0$, we derive the FOC with respect to $t$ as
\begin{align*}
\frac{d\E[v(t)]}{dt} = (r_m-r_m') \left( e^{-2\lambda t} + e^{-\lambda t} \right) -r_f = 0,
\end{align*}
where the SOC trivially holds.
Solving this equation gives the optimal timing of leaving the host country:
\begin{align*}
t = \begin{cases}
0 &\text{if} \ \ r_f \ge 2(r_m -r_m') \\
-\dfrac{1}{\lambda} \log \left( -1 + \dfrac{1}{2}\sqrt{1 + \dfrac{r_f}{r_m-r_m'}} \right) \equiv \widehat{t} &\text{if} \ \ r_f < 2(r_m -r_m')
\end{cases},
\end{align*}
where the term inside the logarithm is smaller than unity if $r_f < 2(r_m -r_m')$.
Especially if the flow return of foreign domestic firms is large enough such that $r_f \ge 2(r_m -r_m')$, multinationals leave the host country at time $s=0$.
This is in a sharp contrast to the case without uncertainty.
Even though the damage of shock is so small that multinationals make a higher flow return than foreign domestic firms after the shock, 
uncertainty itself may lead foreign capital to move out of the host country before the actual shock hits, because (we assume) it expects huge subsequent shocks to come.
If the flow return of foreign domestic firms is small enough such that $r_f < 2(r_m -r_m')$, multinationals stay for a certain period.
The duration becomes shorter when they perceive a higher frequency of disaster occurrence: $d\widehat{t}/d\lambda<0$.

The findings are summarized as follows.

\

\noindent
{\bf Proposition A3 (Disaster risk and the timing of leaving).} \ \ {\it Consider the situation where the host economy is initially at equilibrium $S_1$ and is then hit by a natural disaster at time $T>0$.
The disaster raises the fixed labor input from $F$ to $F' (< F_b)$, where $r_m > r_m'> r_f$ holds.
On the optimal timing of multinational leaving the host country, the following holds:
\begin{itemize}
\item[(i)] If multinationals know the exact timing $T$, they stay in the host country after the disaster and never leave there.
\item[(ii)] If multinationals do not know the exact timing $T$ and only know it follows the cumulative distribution $G(T)=1-e^{-\lambda T}$ and expect huge subsequent disasters to come, they leave the host country at time $s=0$ if $r_f \ge 2(r_m-r_m')$ and do so at time $s=\widehat{t}>0$ otherwise.
\end{itemize}
}

\end{spacing}

\pagebreak

\section*{Data Appendix}

We here provide supplementary tables for empirical examples in Introduction.

\

\begingroup
\begin{spacing}{1.0}
%\begin{adjustwidth}{-1.5cm}{-1.5cm}
\begin{center} 
Table A1. \ Description of variables \
{\small  \begin{tabular}{lcc} \\[-1.8ex] \hline \hline
    Variable & Description & Source  \\ \hline \\[-1.8ex] 
    No. of natural disasters & \multicolumn{1}{>\justify m{7cm}}{Number of natural disasters that record positive financial damages.}  & EM-DAT \\ \hline
    FDI inflows (billion USD) & \multicolumn{1}{>\justify m{7cm}}{FDI net inflows deflated by the price level of real GDP (``\textit{pl$\_$gdpo}'' in PWT, equal to the PPP divided by the nominal exchange rate).}  & WDI \\ \hline
  log (L. GDP) & \multicolumn{1}{>\justify m{7cm}}{Log of real GDP, lagged one year. Real GDP is output-side real GDP at chained PPPs in 2017USD (``\textit{gdppo}'').} & PWT \\ \hline
    L. Tariff ($\%$) & \multicolumn{1}{>\justify m{7cm}}{Average effective tariff rate of manufacturing products weighted by the product import shares corresponding to each partner country, lagged one year. Missing values are linearly interpolated.} & WDI \\ \hline
   L. Unit labor cost & \multicolumn{1}{>\justify m{7cm}}{Nominal labor income divided by real GDP, relative to US value, lagged one year. It is given by the product of the labor share (``\textit{labsh}'') and the price level (``\textit{pl$\_$gdpo}'').}  & PWT  \\ \hline
   \hline \\[-1.8ex] 
  \end{tabular} }
%\end{table}
\end{center}
%\end{adjustwidth}
\noindent {\small \textit{Notes:} EM-DAT: Emergency Events Database by the Centre for Research on the Epidemiology of Disasters, \url{https://www.emdat.be/}} \ \ WDI: World Development Indicators by the World Bank, \url{https://datatopics.worldbank.org/world-development-indicators/} \ \ PWT: Penn World Table 10.0 (\citealp{Feenstraetal2015}), \url{https://www.rug.nl/ggdc/productivity/pwt/?lang=en}
\end{spacing} 
\endgroup

%IMF
%https://data.imf.org/?sk=F8032E80-B36C-43B1-AC26-493C5B1CD33B

%Tax 
%https://taxfoundation.org/publications/corporate-tax-rates-around-the-world/

\pagebreak

\begin{center}
Table A2. \ List of countries \\
\vspace{0.3cm}
\includegraphics[scale=0.85]{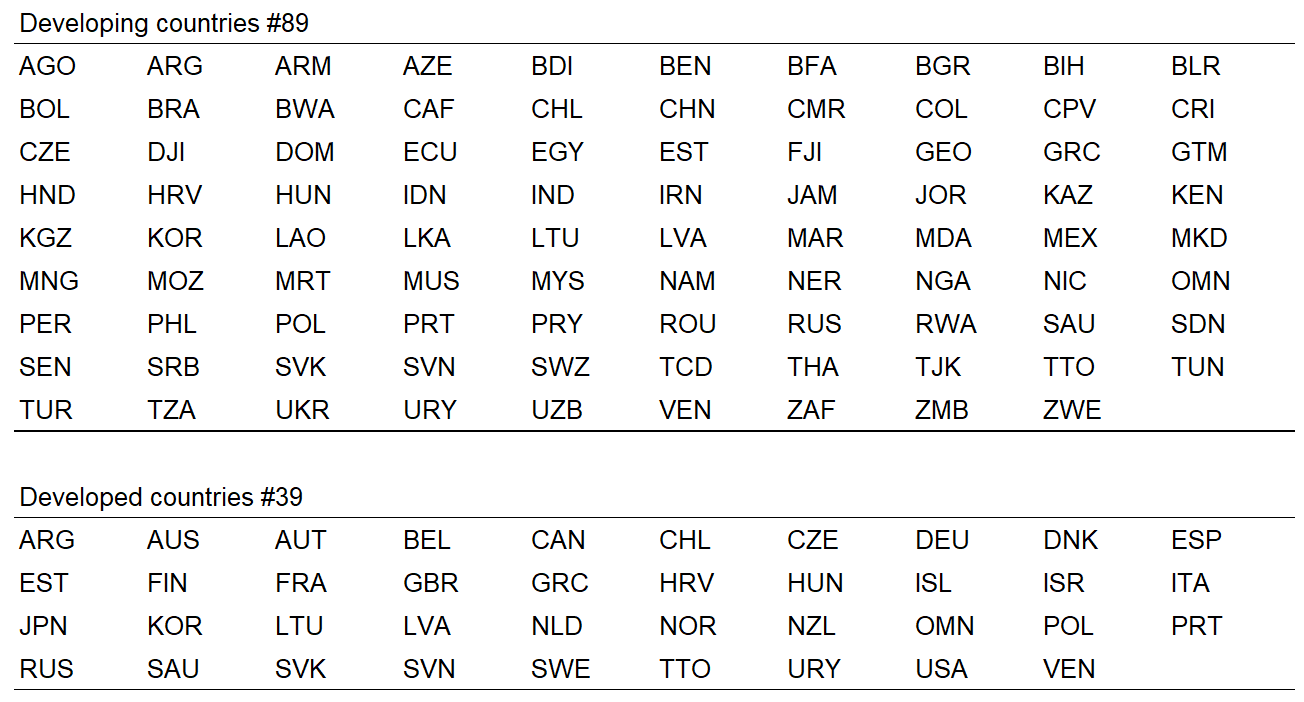}
\end{center}
\vspace{-0.3cm}
\begin{spacing}{1.0}
\noindent {\small \textit{Notes:} Developing countries are those belonging to low-income, lower-middle-income, or upper-middle-income groups at some point in 1991 to 2015, according to the World Bank. See: \url{https://datatopics.worldbank.org/world-development-indicators/the-world-by-income-and-region.html} \ \
Developed countries are those belonging to high-income group at some point in 1991 to 2015.
The list does not include countries identified as tax havens (``big seven'' and ``dots'') by \cite{HinesRice1994}.}
\end{spacing}

\

\begingroup
\begin{center}
Table A3. \ Summary statistics \\
\vspace{0.3cm}
\includegraphics[scale=0.9]{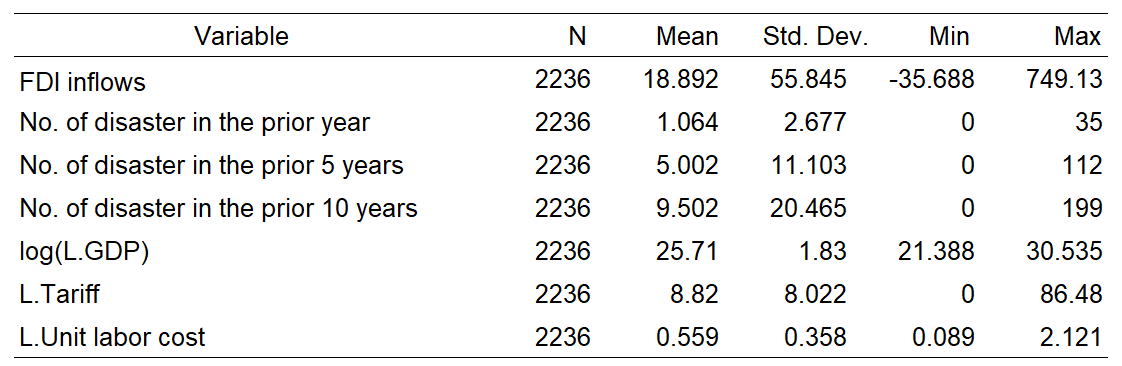}
\end{center}
\endgroup

\

\begingroup
\begin{center}
Table A4. \ Replication of Table 1 without  developing-country dummies \\
\vspace{0.3cm}
\includegraphics[scale=0.9]{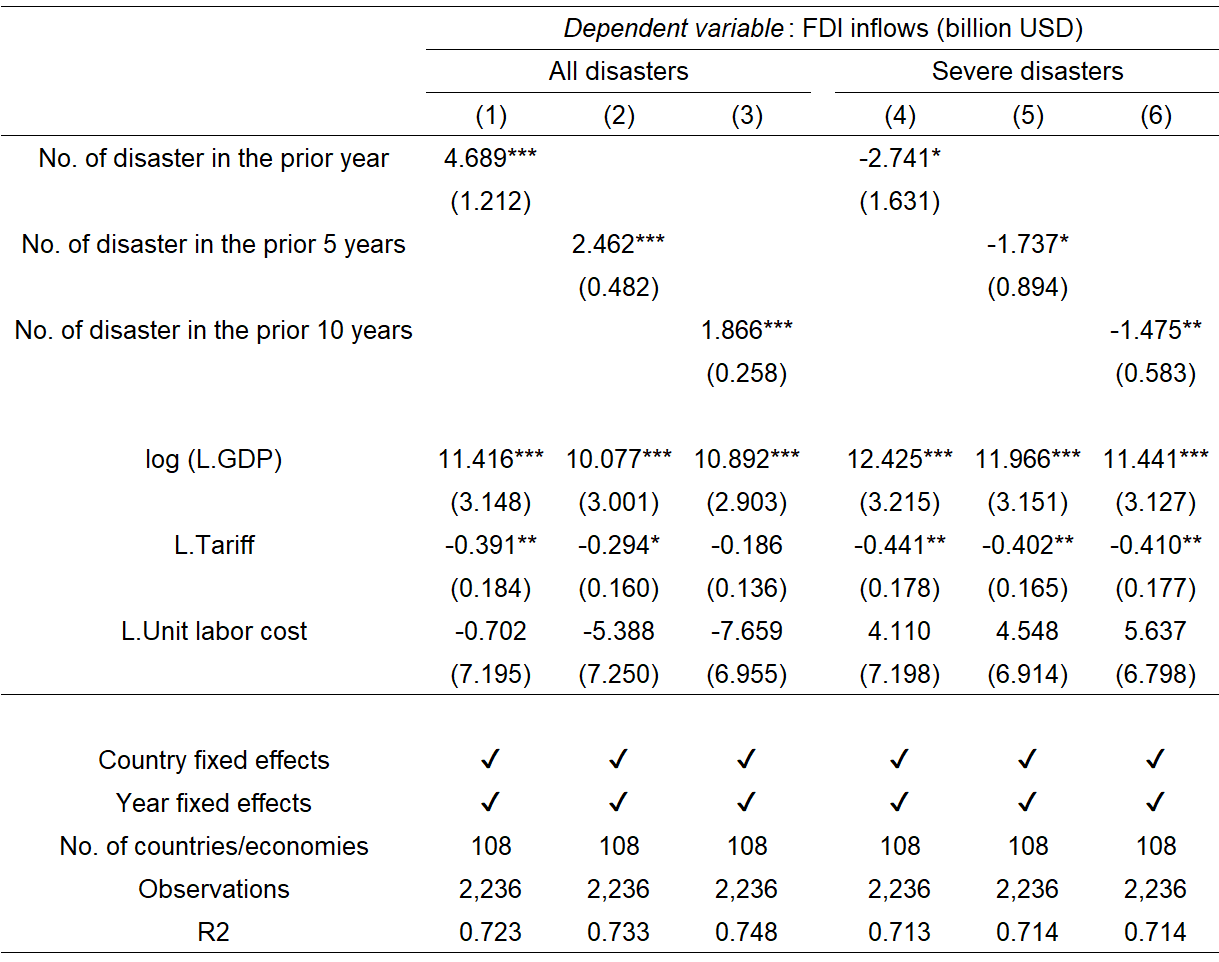}
\end{center}
\vspace{-0.3cm}
\begin{spacing}{1.0}
\noindent {\small \textit{Notes:} Robust standard errors clustered at region$\times$year ($5 \times 25$) are in parentheses.
The sample period is from 1991 to 2015.
FDI inflows (billion USD) are deflated by the price level of real GDP.
A severe disaster in country $i$ in year $t$ is defined as one that records financial damages as a share of GDP exceeding its median for all countries that have ever experienced natural disasters in 1981--2015.
A non-severe disaster in country $i$ in year $t$ is defined as one that is not severe.
The three control variables, the log of real GDP, the average tariff rate, and the unit labor cost, are lagged one year.
We exclude tax haven countries identified by \cite{HinesRice1994}.
\\
$^{*}$Significant at 10$\%$ level; $^{**}$Significant at 5$\%$ level; $^{***}$Significant at 1$\%$ level. }
\end{spacing}
\endgroup

\pagebreak

\begingroup
\begin{center}
Table A5. \ Replication of Table 2 without  developing-country dummies \\
\vspace{0.3cm}
\includegraphics[scale=0.9]{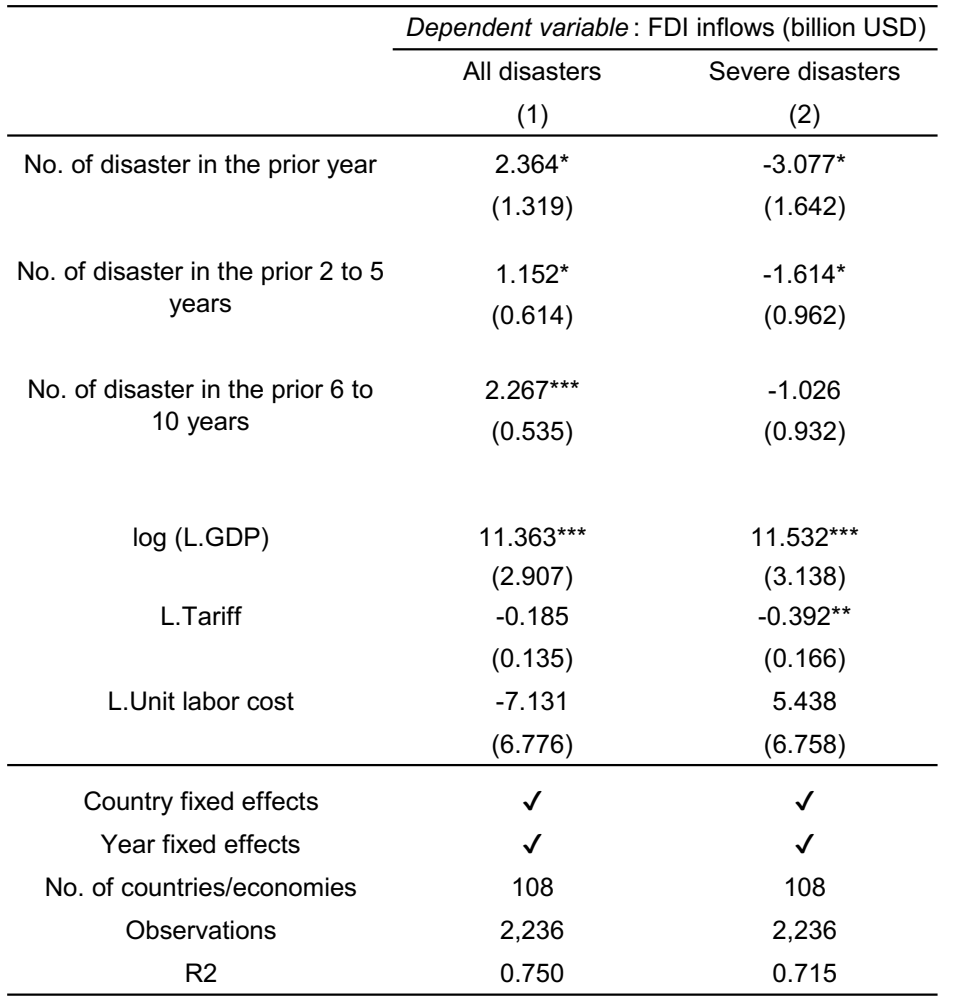}
\end{center}
\vspace{-0.3cm}
\begin{spacing}{1.0}
\noindent {\small \textit{Notes:} Robust standard errors clustered at region$\times$year ($5 \times 25$) are in parentheses.
The sample period is from 1991 to 2015.
FDI inflows (billion USD) are deflated by the price level of real GDP.
A severe disaster in country $i$ in year $t$ is defined as one that records financial damages as a share of GDP exceeding its median for all countries that have ever experienced natural disasters in 1981--2015.
A non-severe disaster in country $i$ in year $t$ is defined as one that is not severe.
The three control variables, the log of real GDP, the average tariff rate, and the unit labor cost, are lagged one year.
We exclude tax haven countries identified by \cite{HinesRice1994}.
\\
$^{*}$Significant at 10$\%$ level; $^{**}$Significant at 5$\%$ level; $^{***}$Significant at 1$\%$ level. }
\end{spacing}
\endgroup

\pagebreak

\

\begin{spacing}{1.0}
\bibliographystyle{C:/Users/Hayato/Dropbox/apalike.bst}
\bibliography{C:/Users/Hayato/Dropbox/TeX/texlive/2019/texmf-dist/bibtex/econ_ref_hayato}
\end{spacing}

\end{document}